\documentclass[aps,preprint,prd,epsfig]{revtex4}
\usepackage{amssymb}
\usepackage{amsmath}
\usepackage{graphicx}
\usepackage{fancyhdr}
\newcommand{\be}{\begin{equation}}
\newcommand{\ee}{\end{equation}}
\newcommand{\bea}{\begin{eqnarray}}
\newcommand{\eea}{\end{eqnarray}}
\newcommand{\bes}{\begin{subequations}}
\newcommand{\ees}{\end{subequations}}

\newcommand{\bc}{\begin{center}}
\newcommand{\ec}{\end{center}}
\usepackage{amsmath, amssymb} 
\usepackage{slashed}

\begin{document}

\title{ Inverse type II seesaw mechanism  and its signature at the LHC and ILC }

\author{F. F. Freitas}
\affiliation{{ Departamento de
F\'{\i}sica, Universidade Federal da Para\'\i ba, Caixa Postal 5008, 58051-970,
Jo\~ao Pessoa, PB, Brasil}}

\author{C. A. de S. Pires}
\affiliation{{ Departamento de
F\'{\i}sica, Universidade Federal da Para\'\i ba, Caixa Postal 5008, 58051-970,
Jo\~ao Pessoa, PB, Brasil}}

\author{P. S. Rodrigues da Silva}
\affiliation{{ Departamento de
F\'{\i}sica, Universidade Federal da Para\'\i ba, Caixa Postal 5008, 58051-970,
Jo\~ao Pessoa, PB, Brasil}}
\date{\today}

\begin{abstract}
The advent of the LHC, and the proposal of building  future colliders as the ILC, both  programmed to explore new physics at the TeV scale, justifies the recent interest in studying  all kind of seesaw mechanisms  whose signature lies on such energy scale. The natural candidate for this kind of seesaw mechanism is the inverse one. The conventional inverse seesaw mechanism is implemented in an arrangement  involving  six new heavy neutrinos in addition to the three standard ones. In this paper we develop the inverse seesaw mechanism  based on Higgs triplet model and probe its signature  at the LHC and  ILC.  We argue that the conjoint analysis of the LHC together with the  ILC may confirm the mechanism and, perhaps, infers the hierarchy of the neutrino masses.
\\
%
\end{abstract}

\maketitle

\section{Introduction}

The essence of inverse seesaw (ISS) mechanisms rest upon the assumptions that  standard neutrinos are Majorana particles and that lepton number is explicitly violated at low energy scale\cite{ISSI}. Its implementation  into the standard model (SM)  demands  new fermions, in the singlet\cite{ISSIII} or triplet form\cite{ISSII}, or  new Higgs, in the triplet form\cite{ISSII}. In this paper we are interested in the ISS mechanism implemented in the framework of Higgs triplet model.

Higgs triplet model\cite{TPCheng} is very versatile in implementing seesaw mechanisms. If we admit lepton number is explicitly violated at high energy scale, the Higgs triplet model provides the conventional seesaw mechanism\cite{seesawII}, but if we admit the contrary, the Higgs triplet model provides the inverse seesaw mechanism\cite{ISSII}.  The signature of the former case is out of the range of current accelerators since it phenomenology is manifested at GUT scale. In the latter case,  the signature may manifest from electroweak until TeV scale   and then may be probed   at the LHC or at  future TeV  colliders as ILC.  

In general, seesaw mechanisms require that  neutrinos  be  Majorana particles\cite{seesaw, seesawII,seesawIII}. Thus their implementations demand explicit violation of the lepton number. In conceptual level, what distinguish conventional from inverse seesaw mechanisms is that in the former  lepton number is explicitly violated at very high energy scale, while in the  latter lepton number is violated explicitly at low energy scale. This is the reason for the term ''inverse''.

Higgs triplet model consists in adding, to the standard model,  a  triplet of scalars,  $\Delta$,   having  hypercharge, $Y=2$, and lepton number, $L=-2$\cite{TPCheng}. In this model lepton number may be explicitly violated in the potential of the model through the trilinear term, $\mu \Phi^T  \Delta \Phi$, where $\Phi$ is the standard Higgs doublet and $\mu$ is an energy parameter.  When we admit  that  $\mu$ lies around the keV scale, the Higgs triplet model recovers the inverse seesaw mechanism\cite{ISSII}.   We refer to this case as the inverse type II seesaw (ISSII) mechanism. This mechanism  was first  discussed in the literature in the Refs. \cite{ISSII} and since then has received  few attention. While the conventional  ISS mechanism  provides  tiny neutrino masses\cite{ISSI},  the ISSII mechanism provides tiny vacuum expectation value for the neutral component of $\Delta$\cite{ISSII}. The main signature of the ISSII mechanism is doubly   charged  scalars with masses around TeV. In this work we  revisit such mechanism and probe its signature in the form of  doubly charged scalars at the LHC and  ILC for scenarios with neutrinos masses obeying normal and inverted hierarchies.

This work is organized as follows: in Sec.~\ref{sec2} we review the  inverse seesaw  mechanism  in order to establish the framework for the ISSII  to be developed in Sec.~\ref{sec3}. We then work out the mass spectrum of the scalar sector of the model in Sec.~\ref{sec4}. Next, in Sec.~\ref{sec5}, we pursue the phenomenological bounds concerning the rare lepton decay $\mu\rightarrow e + \gamma$ and in Sec.~\ref{sec6} we study the collider signature of the ISSII at the ILC and LHC. We present our concluding remarks in Sec.~\ref{sec7}.

\section{Inverse seesaw mechanisms}
\label{sec2}
In  ISS mechanisms   lepton number is postulated to be explicitly  violated at low energy scale and neutrinos gain the mass expression  $M_\nu = m_ D^2 \mu M_N^{-2}$ \cite{ISSI,ISSIII,ISSII}. Their signatures are new scalars or fermions with mass around TeV. 

 For illustrative reasons, we present the main ingredients of the conventional ISS mechanism\cite{ISSI}. It  is based on the extension of the SM  by six new  singlet  neutrinos ($N_{i}\,,\,S_{i}$) with $i=1,2,3$ . The mechanism is engendered by  the following mass terms,
\begin{equation}
{\cal L}=-\bar \nu m_D N  - \bar N M_N S- \frac{1}{2} \bar S^C \mu S + {\mbox h.c.}\,,
\label{massterms}
\end{equation}
where $\nu=(\nu_e\,,\,\nu_\mu \,,\, \nu_\tau )$ are the standard neutrinos in the flavor basis. These mass terms may be expressed  in the matrix form
\begin{equation}
M_\nu=
\begin{pmatrix}
0 & m^T_D & 0 \\
m_D & 0 & M_N^T\\
0 & M_N & \mu
\end{pmatrix},
\label{ISSmatrix}
\end{equation}
whose basis is $(\nu\,,\,N\,,\, S)$,with $m_D$, $M_N$  and $\mu$ being  $3\times 3$ mass matrices. Without loss of generality,  we consider that $\mu$ is diagonal, and suppose the following hierarchy , $\mu \ll m_D \ll M_N$. What makes the texture  in Eq.~(\ref{ISSmatrix}) interesting from the phenomenological point of view is that after block diagonalization of $M_\nu$, we obtain, in a first approximation, the following effective neutrino mass matrix for the standard neutrinos\cite{rodehojam}:
\begin{equation}
m_\nu = m_D^T M_N^{-1}\mu (M_N^T)^{-1} m_D.
\label{inverseseesaw}
\end{equation}
Assuming $m_D =Y_D v_\phi$, $M_N =Y_N M$ and $\mu$ diagonal, we have,
 \begin{equation}
m_\nu =Y \frac{v^2_\phi}{M^2} \mu  ,
\label{issexpression}
\end{equation}
where $Y=Y_D^T Y_N^{-1} (Y_N^T)^{-1} Y_D $.

In  ISS mechanisms the magnitude of the standard neutrino masses is dictated by the expression $m_\nu \approx \frac{v^2_\phi}{ M^{2}} \mu$. Thus, for $v_\phi$ at electroweak scale, $M$ at TeV scale requires $\mu$ around few keV\cite{explanation,realization}.  The signature of the conventional ISS mechanism are six heavy neutrinos,  $N_i$ and $S_i$, and they can be probed at the LHC or at the future ILC. Bounds on the conventional ISS mechanism are obtained from rare leptonic decays mediated by the heavy neutrinos and non-unitarity effects caused by the mixing of the heavy neutrinos with the standard ones. 

ISS mechanisms may be engendered, too,  through  new fermion or new  Higgs in the triplet form.  These cases have received few attention in the literature.  In what follow we discuss the implementation and the phenomenology of the ISS mechanism engendered through Higgs triplet. 


\section{The inverse type II  seesaw mechanism.}
\label{sec3}
The implementation of  what we call the inverse type II seesaw (ISSII) mechanism requires we add to the SM the Higgs triplet
\begin{equation}
\Delta\equiv \left(\begin{array}{cc}
\frac{\delta^{+}}{\sqrt{2}} & \delta^{++} \\ 
\delta^{0} & \frac{-\delta^{+}}{\sqrt{2}}
\end{array} \right)\,\sim (1, 3,2), 
\end{equation}
which, together with the SM scalar doublet, $\Phi = (\phi^{+}\,\,\,\, 
\phi^{0})^T\sim(1,2,-1)$, compose the following potential,
\begin{eqnarray}
V(\Phi,\Delta) &=& -m_{H}^{2}\Phi^{\dagger}\Phi + \frac{\lambda}{4}(\Phi^{\dagger}\Phi)^{2} + M^{2}_{\Delta}Tr[(\Delta^{\dagger}\Delta)]+[\mu(\Phi^T i\sigma^{2}\Delta^{\dagger}\Phi)+H.c]\nonumber \\
&&+\lambda_{1}(\Phi^{\dagger}\Phi)Tr[(\Delta^{\dagger}\Delta)] +\lambda_{2}(Tr[(\Delta^{\dagger}\Delta)])^{2} +\lambda_{3}Tr[(\Delta^{\dagger}\Delta)^{2}] \nonumber \\
&&+ \lambda_{4}\Phi^{\dagger}\Delta^{\dagger}\Delta\Phi + \lambda_{5}\Phi^{\dagger}\Delta\Delta^{\dagger}\Phi\,.
\label{potential}
\end{eqnarray}
Assuming  that $\Delta$ carries lepton number, we then have that the trilinear term in the  potential above violates  lepton number explicitly.

In order to develop the scalar sector  of this model and obtain its scalar spectrum, it is necessary to shift  the neutral components of $\Phi$ and  $\Delta$ in the conventional way
\begin{eqnarray}
 \phi^{0} , \delta^{0} \rightarrow \frac{1}{\sqrt{2}}\left(  v_{\phi , \Delta} 
+R_{ _{\phi ,\Delta} } +iI_{_{\phi ,\Delta} }\right) \,,
\label{vacua} 
\end{eqnarray}
where $v_\phi$ and $v_\Delta$ are the  vacuum expectation values (VEV) of the fields $\phi^0$ and $\delta^0$, respectively. The VEV $v_\Delta$ modifies softly the $\rho$-parameter in the following way: $\rho=\frac{1+\frac{2v^2_\Delta}{v^2_\phi}}{1+\frac{4v^2_\Delta}{v^2_\phi}}$. The current value $\rho=1.0004^{+0.0003}_{-0.0004}$\cite{PDG} implies the following upper bound $v_\Delta < 5$GeV. The regime of energy for $v_\Delta$ we are interested in  lies around few eVs, which satisfies the upper bound put by the $\rho$ parameter.

After the shift above, on imposing the minimum conditions over the potential, we obtain a set of constraint equations over the parameters of the potential,
\begin{eqnarray}
&&-m^{2}_{H}+\frac{1}{4}(v^{2}_{\phi}\lambda +2v_{\Delta}(v_{\phi}(\lambda_{1}+\lambda_{4})-2\sqrt{2}\mu))=0 ,\nonumber \\
&&M^{2}_{\Delta} v_{\Delta}+v^{3}_{\Delta}(\lambda_{2}+\lambda_{3})+\frac{v^{2}_{\phi}v_{\Delta}}{2}(\lambda_{1}+\lambda_{4}) -\frac{1}{\sqrt{2}}v^{2}_{\phi}\mu=0\,.
\label{constraints}
\end{eqnarray}
The versatile of the Higgs triplet model concerning seesaw mechanisms arise now. Note that the second constraint above provides
\begin{equation}
 v_{\Delta}\simeq \frac{1}{\sqrt{2}}v_\phi M^{-1}_\Delta\mu v_\phi M^{-1}_\Delta.
 \label{ISSequation}
\end{equation}
Perceive that, on assuming that lepton number is  explicitly  broken at high energy scale, $M$, on taking $\mu=M_\Delta = M$ , Eq. (\ref{ISSequation}) provides
\begin{equation}
v_{\Delta}\simeq \frac{1}{\sqrt{2}}\frac{v^2_\phi}{ M}.
\label{typeIISS}
\end{equation}
This is the well known type II seesaw mechanism for $v_\Delta$.  For $M$ at GUT scale, namely $M \sim 10^{14}$ GeV and $v_\phi =10^2$ GeV, we have $v_\Delta$ around eV scale.

 On the contrary, on assuming that  lepton number is explicitly  broken  at low energy scale, namely, that   $\mu$  is much smaller that $M_\Delta$ and $v_\phi$, Eq. (\ref{ISSequation}) provides
 \begin{equation}
  v_\Delta  \simeq \frac{1}{\sqrt{2}} \frac{v^2_\phi}{M^2_\Delta}\mu .
  \label{ISSII}
  \end{equation}

This is the inverse type II seesaw mechanism for the VEV $v_\Delta$.  Note that   $\mu$ at the keV scale requires $M_\Delta$ as high as TeV scale for providing  $v_\Delta$ at eV scale (for $v_\phi$ is the electroweak scale).  From now on we restrict our investigation  to the development of the Higgs triplet model in the regime of energy that promote the realization of the ISSII mechanism. 

We finish this section obtaining the expression for the neutrino masses provided by the Higgs triplet model. The Yukawa interactions involving $\Delta$ and the standard lepton doublet  $L=(\nu \,\,,\,\, e)_L^T$ are
\begin{equation}
\mathcal{L}_{Y}=Y_{ij}\bar{L}^{c}_{i} i \sigma_{2}\Delta L_{j} + H.c. 
\label{Yukawacoupling}
\end{equation}
When $\Delta$ develops  VEV, we obtain the following general neutrino mass expression
\begin{equation}
 m_\nu=\frac{Y}{\sqrt{2}} v_{\Delta}.
 \label{lISSIImasssexression}
 \end{equation}
  Substituting the expression for $v_\Delta$ given in Eq. (\ref{ISSII}), we have
\begin{equation}
m_\nu=\frac{Y}{2}\frac{v^2_\phi}{ M^{2}_\Delta}\mu .
\label{neutrinomassISSII}
\end{equation}
Observe that the above expression for the neutrino masses recovers the one that appears in the ISS mechanism in Eq. (\ref{issexpression}). 

The advantages of the ISSII mechanism compared to the other two scenarios of ISS mechanisms\cite{ISSI,ISSIII} are twofold: it does not  modify the neutrino sector, that is,  the neutrinos of the model are the standard ones.  It provides a clear phenomenology in the form of electrically  charged scalars with masses around TeVs which may be probed mainly at the ILC and perhaps at the LHC. 

The goal of this work is to revisit the ISSII mechanism and  probe its phenomenology  in accelerators as LHC and ILC. This requires we have in hand the spectrum of scalars of the model. Thus, in the next section we obtain the spectrum of scalars of the Higgs triplet model in the regime of energy that trigger the ISSII mechanism.

\section{Spectrum of scalars}
\label{sec4}

Motivated by the running of the LHC at TeV scale, the spectrum of scalars composing the triplet $\Delta$ has been extensively investigated in the last years\cite{spectrumDelta}.  It is important to stress that in order to pursue such investigations the focus had to be on the parameter space appropriate to leave some track of the new scalars in LHC. This means that, if the aim is to look for these new scalars through enhanced couplings to SM particles, suitable for LHC searches, $v_\Delta$ has invariably to be taken far away from the eV scale. Precisely $v_\Delta \geq 10^{-4}$ GeV.  The price to be paid rests on the loss of any natural explanation for the smallness of neutrino masses since this requires  very tiny Yukawa couplings in order to have neutrino masses that conciliate atmospheric and solar oscillation.  

In the opposite direction, in this paper we give emphasis  on the parameter space of the Higgs triplet model that trigger the  ISSII mechanism,  which means to obtain the scalar spectrum of the model for a scenario where $v_\Delta$ and $\mu$ are  kept small enough lying in the range from eV to keV. Here we consider the consequence of this choice for the parameters in the scalar spectrum of ISSII so as to explore its implications in the next section.

From the scalar  potential, Eq.~(\ref{potential}), together with the constraint equations, Eq.~(\ref{constraints}), we obtain  the following mass matrix  for the CP-even  neutral scalars in the basis $(R_\phi\,,\,R_\Delta)$,
\begin{equation}
m^{2}_{h}=\left(\begin{array}{cc}
A & B \\ 
B & C
\end{array} \right),
\label{cpevenmassmatrix}
\end{equation}
where the terms $A$, $B$ and $C$ are,
\begin{equation}
\begin{split}
A&=\frac{v^{2}_{\phi}\lambda}{2} ,\\
B&=\frac{v_{\phi}(v_{\Delta}(\lambda_{1}+\lambda_{4})-\sqrt{2}\mu)}{2v_{\Delta}} ,\\
C&=\frac{4v^{3}_{\Delta}(\lambda_{2}+\lambda_{3})+\sqrt{2}v^{2}_{\phi}\mu}{2v_{\Delta}}.
\end{split}
\label{ABC}
\end{equation}
In the limit $v_{\phi}\gg \mu ,  v_\Delta$, we obtain the following eigenvalues,
\begin{equation}
\begin{split}
m^{2}_{h^{0}}&\simeq \frac{v^{2}_{\phi}\lambda}{4},\\
m^{2}_{H^{0}}&\simeq m^{2}_{h^{0}}+\left( \frac{1}{\sqrt{2}}\frac{\mu}{v_{\Delta}}\right)v^{2}_{\phi}.
\end{split}
\label{cpevenmass}
\end{equation}
Regarding the eigenvectors, we obtain, 
\begin{equation}
\left(\begin{array}{c}
h^{0} \\ 
H^{0}
\end{array}\right)\simeq \left(\begin{array}{cc}
1 & \sqrt{\frac{v_{\Delta}}{v_{\phi}}} \\ 
- \sqrt{\frac{v_{\Delta}}{v_{\phi}}} & 1
\end{array}\right) \left(\begin{array}{c}
R_{\phi} \\ 
R_{\Delta}
\end{array}\right) .
\end{equation}

We recognize that $h^0$ is the standard Higgs, while $H^0$ is a second Higgs that survives in the model. For $v_\Delta \approx 1$eV and $v_\phi \approx 10^2$GeV, we get $\frac{v_\Delta}{v_\phi}\approx 10^{-11}$.  In this case we see that  $h^0$ decouples from $H^0$. 

For the CP-odd neutral scalars we get the mass matrix in the basis $(I_\Delta\,,\,I_\phi)$, 
\begin{equation}
m^{2}_{A}=\sqrt{2}\mu\left(\begin{array}{cc}
2v_{\Delta} & -v_{\phi} \\ 
-v_{\phi} & \frac{v^{2}_{\phi}}{2v_{\Delta}}
\end{array}  \right).
\label{cpoddmassmatrix}
\end{equation}
In the limit $v_{\phi}\ggg \mu > v_{\Delta}$,  we obtain the following eigenvalues,
\begin{equation}
\begin{split}
m^{2}_{G^{0}}&=0, \\
m^{2}_{A^{0}}&\simeq \frac{1}{\sqrt{2}}v^{2}_{\phi}\frac{\mu}{v_{\Delta}},
\end{split}
\end{equation}
with their respective eigenvectors, 
\begin{equation}
\left(\begin{array}{c}
G^{0} \\ 
A^{0}
\end{array}\right)=\left(\begin{array}{cc}
\cos\beta & \sin\beta \\ 
-\sin\beta & \cos\beta
\end{array}\right)\left(\begin{array}{c}
I_{\phi} \\ 
I_{\Delta}
\end{array}\right),
\label{mixscalars}
\end{equation}
where,
\begin{equation}\label{eq18}
\sin\beta =\frac{2v_{\Delta}}{\sqrt{v^{2}_{\phi}+2v^{2}_{\Delta}}}\,\,\,,\,\,\,
\cos\beta  =\frac{v_{\phi}}{\sqrt{v^{2}_{\phi}+2v^{2}_{\Delta}}}.
\end{equation}

As we are assuming $v_{\phi}\ggg  v_\Delta$, we have that $\sin\beta \rightarrow 0$ and $\cos\beta \rightarrow 1$ which means that $G^0$ decouples from $A^0$. Also, $G^0$ is a Goldstone boson that will be eaten by the SM neutral gauge boson $Z$,  and $A^0 $ is a massive CP-odd scalar that survives in the particle spectrum.

The mass matrix for the singly charged scalars in the basis $(\delta^+ \,,\, \phi^+)$ is given by,
\begin{equation}
m^{2}_{+}=(\sqrt{2}\mu-\frac{v_{\Delta}\lambda_{4}}{2})\left(\begin{array}{cc}
v_{\Delta} & -\frac{v_{\phi}}{\sqrt{2}} \\ 
-\frac{v_{\phi}}{\sqrt{2}} & \frac{v^{2}_{\phi}}{2\sqrt{2}}
\end{array}\right) 
\end{equation}
In the limit $v_{\phi}\ggg v_{\Delta}, \mu$,  we obtain the following  eigenvalues,
\begin{equation}
\begin{split}
& m^{2}_{G^{+}}=0, \\
& m^{2}_{H^{+}}\simeq \frac{\sqrt{2}}{2}(\frac{\mu}{v_{\Delta}}-\lambda_{4})v^{2}_{\phi},
\end{split}
\label{hpmass}
\end{equation}
where $G^{+}$ is the Goldstone boson eaten by the SM charged gauge bosons, $W^\pm$, while $H^\pm $ are massive scalars remaining in the spectrum. The mixing mass matrix for $\delta^\pm$ and  $\phi^\pm$ is the same one that appears in Eq.~(\ref{mixscalars}). Thus, the singly charged scalars decouple too.

In regard to the doubly charged scalars,  $\delta^{\pm\pm}$,  we obtain the following expression for its mass in the limit $v_{\phi}\ggg \mu , v_{\Delta} $,
\begin{equation}
m^{2}_{\delta^{++}}\simeq M^2_\Delta-\lambda_{4}v^{2}_{\phi}.
\label{hppmass}
\end{equation}

All this review has the intention of recalling the noticeable fact that in the ISSII mechanisms, the new scalars composing the triplet $\Delta$ decouple from the SM scalars. In other words, we have the Englert-Brout-Higgs boson, $h^0$, and the new massive scalars, $H^0$, $A^0$, $H^\pm$ and $\delta^{\pm\pm}$, all decoupled in the particle spectrum. The difference is that  in the ISSII mechanism, $v_\Delta$ around eV requires $M_\Delta$ at most at TeV scale. The degeneracy among  $H^+$ and $\delta^{++}$ is a consequence of taking $v_\Delta$ at eV scale. 

 However, because of the decoupling of the new scalar from the standard ones, they do not couple with quarks. Thus their probe at the LHC is prompted by their couplings with the gauge bosons. For this case, as we said before, the probe at the LHC is favored only for $v_\Delta \geq 10^{-4}$ GeV, which is not the case of the ISSII mechanism.

 We argue in this work that the fairest place to probe for the ISSII mechanism is at the ILC, but before discussing this we first delve into the constraints coming from the rare decay $\mu \rightarrow e\gamma$, which must impose some restriction to the parameter space we are interested in.

\section{Neutrino masses and the rare lepton decay $\mu \rightarrow e\gamma$}
\label{sec5}
The neutrino mass matrix  in the flavor basis,   given in Eq. (\ref{neutrinomassISSII}),   is related to the physical mass matrix, $m^D_\nu$,   through a  $3\times 3$ unitarity mixing matrix $U$ in the following way
\begin{equation}\label{eq31}
m^D_{\nu}=U m_{\nu} U^{\dagger},
\end{equation}
where  $m^{D}_{\nu}=\mbox{diag}(m_1\,,\,m_2\,,\,m_3)$  and  $U$ is the neutrino mixing matrix, which may be   parametrized in the general way by
\begin{equation}\label{eq32}
U=\left(\begin{array}{ccc}
c_{12}c_{13} & s_{12}c_{13} & s_{13} \\ 
-s_{12}c_{23}-c_{12}s_{23}s_{13} & c_{12}c_{23}-s_{12}s_{23}s_{13} & s_{23}c_{13} \\ 
s_{12}s_{23}-c_{12}c_{23}s_{13} & -c_{12}s_{23}-s_{12}c_{23}s_{13} & c_{23}c_{13}
\end{array}\right) \,,
\end{equation}
where $c_{ij}=\cos\theta_{ij}$ and  $s_{ij}=\sin\theta_{ij}$, while in this work we neglect CP violation phases. Thus, on  inverting  the Eq. (\ref{eq31}) we obtain
\begin{equation}\label{eq31inverted}
m_{\nu}=U^{\dagger} m^D_{\nu} U,
\end{equation}
Combining equations   (\ref{neutrinomassISSII})  and (\ref{eq31inverted}), we obtain
\begin{equation}\label{yukawa}
Y_{ij}=\frac{1}{v_\Delta} U^{\dagger}_{ik} m^D_{{\nu}_{kk}} U_{kj},
\end{equation}
In opening this relation, we get the expressions for the Yukawa entries
\begin{eqnarray}
&&Y_{11}=\frac{\sqrt{2}}{v_{\Delta}}(c^{2}_{12}(m_{1}c^{2}_{13}+m_{2}s^{2}_{12})+m_{3}s^{2}_{13}),\nonumber \\
&& Y_{22}=\frac{\sqrt{2}}{v_{\Delta}}(m_{3}c^{2}_{13}s^{2}_{23} +m_{1}(c_{23}s_{12}+c_{12}s_{13}s_{23})^{2} + m_{2}(c_{12}c_{23}-s_{12}s_{13}s_{23})^{2}),\nonumber \\
&&Y_{33}=\frac{\sqrt{2}}{v_{\Delta}}(m_{3}c^{2}_{13}c^{2}_{23} +m_{2}(c_{23}s_{12}s_{13}-c_{12}s_{23})^{2} + m_{1}(c_{12}c_{23}s_{13}-s_{12}s_{23})^{2}), \nonumber \\
&&Y_{12}=\frac{\sqrt{2}}{v_{\Delta}}(c_{12}(m_{2}c_{12}-m_{1}c_{13})c_{23}s_{12} 
+((m_{3}-m_{1}c^{2}_{12})c_{13}-m_{2}c_{12}s^{2}_{12})s_{13}s_{23}), \nonumber \\
&&Y_{13}=\frac{\sqrt{2}}{v_{\Delta}}(c_{23}s_{13}((m_{3}-m_{1}c^{2}_{12})c_{13} -m_{2}c_{12}s^{2}_{12})+c_{12}s_{12}s_{23}(m_{2}c_{12}+m_{1}c_{13})),\nonumber \\
&&Y_{23}=\frac{\sqrt{2}}{v_{\Delta}}(-\frac{1}{2}(m_{2}-m_{1}(1-2s_{23}^2).(2s_{12}c_{12}))2s_{12}c_{12}s_{13}\nonumber \\
&&+c_{23}s_{23}(m_{3}c^{2}_{13}+ c^{2}_{12}(m_{2}+m_{1}s^{2}_{13})+s^{2}_{12}(-m_{1}+m_{2}s^{2}_{13}))).
\label{yukawarelations}
\end{eqnarray}

According to recent data on neutrino physics, the values of the angles involved in the above mixing matrix are\cite{fogli},
\be
\theta_{12}\simeq \frac{\pi}{5.4}\,\,\,\,,\,\,\,\theta_{23}\simeq \frac{\pi}{4}\,\,\,\,,\,\,\,\,\theta_{13}\simeq \frac{\pi}{20},
\label{anglesdata}
\ee  
while for  the masses of the  neutrinos,  for the normal hierarchy (NH) and inverted hierarchy (IH) cases, we have
\bea
&&m_{1}\,\,\,,  m_{2}=\sqrt{m_{1}^{2} + \Delta m_{\odot}^{2}}\,\,\,, m_{3}=\sqrt{m_{1}^{2}+\Delta m_{atm}^{2}} \,\,\,(\mbox{NH}),\nonumber \\
&&m_3\,\,\,, m_{1}=\sqrt{m_{3}^{2}+\Delta m_{atm}^{2}-\Delta m_{\odot}^{2}},  m_{2}=\sqrt{m_{3}^{2} + \Delta m_{atm}^{2}} \,\,\,(\mbox{IH}),
\label{hierarchy}
\eea
with $\Delta m_{\odot}^{2}\simeq 0.0086 $~eV$^2$ and  $\Delta m_{atm}^{2}\simeq 0.048$~eV$^2$.

The Higgs triplet  $\Delta$  leads to rare leptonic decays mediated by the charged scalars $H^+$ and $\delta^{++}$ engendered by the Eq.~(\ref{Yukawacoupling}). The more stringent of the rare decays is  $\mu^-\rightarrow e^-+\gamma$ which occurs through a loop mediated by the charged scalars $\delta^{\pm\pm}$ and $H^+$. For the case of degeneracy among $H^+$ and $\delta^{\pm \pm}$,  the branching ratio (BR) for this process is given by \cite{raredecay},
\begin{equation}
BR(\mu\rightarrow e\gamma)\simeq \frac{27\alpha\vert Y_{11}Y_{12} + Y_{12}Y_{22} + Y_{13}Y_{32}\vert^{2}}{64\pi G^{2}_{F}m^{4}_{\delta^{++}}}\,,
\label{mudecay}
\end{equation}
where $\alpha$ is the fine structure constant, $Y_{ij}$ are  the Yukawa coupling constants given in Eq.~(\ref{yukawarelations}), $G_{F}$ is the fermi constant and $m_{\delta^{\pm\pm}}$ is the mass of doubly charged scalar. The current experimental bounds on this process is  $BR(\mu\rightarrow e\gamma)< 5.7 \times 10^{-13}$\cite{MEG}. 
Our proposal here is to check all possible values for  $M_\Delta$ that lead to a $v_\Delta$ at the eV scale and concomitantly obey the bound $BR(\mu\rightarrow e\gamma)< 5.7 \times 10^{-13}$.  

To achieve our proposal  we substitute  the experimental values of the mixing angles $\theta_{12}$, $\theta_{23}$ and $\theta_{13}$  in  Eq. (\ref{yukawarelations}),  fix the values  of $m_1$,   $m_2$ and $m_3$ through    $\Delta m_{\odot}^{2}$ and $\Delta m_{atm}^{2}$ for the cases of NH ( where we take $m_1=0$), and IH ( where we take $m_3=0$) and  substitute them  in Eq. (\ref{yukawarelations}). In this way all the Yukawa couplings get  depending of $v_\Delta$, only. We use such Yukawa couplings throughout  this paper. We must always keep in mind that  $v_\Delta$, $\mu$ and $M_\Delta$ are related to each other through  Eq. (\ref{ISSequation}).   

 On substituting these Yukawa couplings  in the expression for $BR(\mu\rightarrow e\gamma)$ given in Eq. (\ref{mudecay}) we have that  bounds from $BR(\mu\rightarrow e\gamma)$ results in bounds on $M_\Delta$ and $v_\Delta$.
\begin{figure}[tb]
\begin{center}
\includegraphics[scale=0.46]{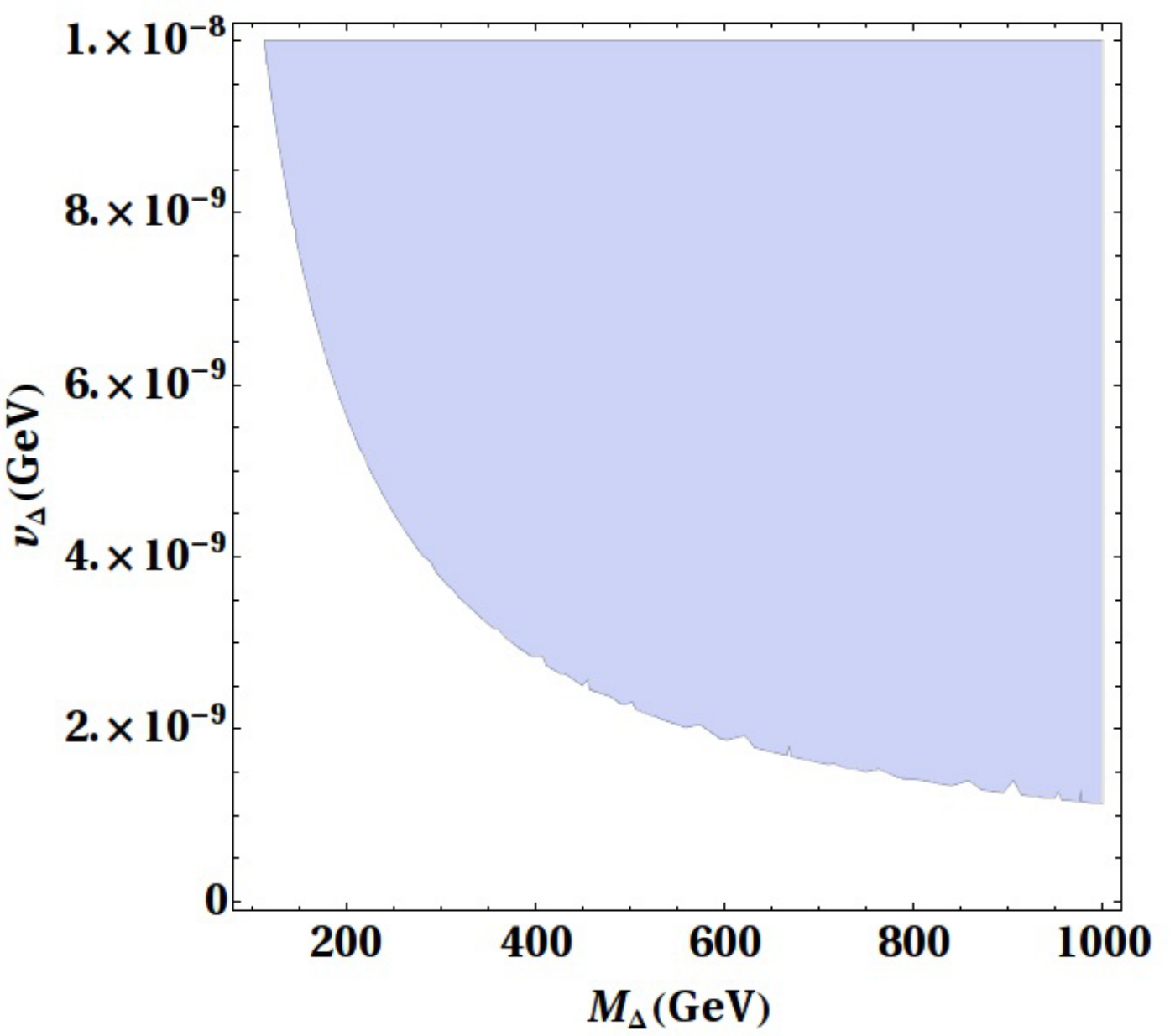}
\caption{ Region in the parameter space  for the case of NH that conforms to the bound  $BR(\mu\rightarrow e\gamma)< 5.7\times 10^{-13}$. The shaded area is the allowed region.}
\label{fig1}\end{center}
\end{figure}

In FIG.~\ref{fig1} we present our result for the  NH case, only.  The plot show  the values of $M_\Delta$ and $v_\Delta$ constrained to obey the upper bound on the rare $\mu\rightarrow e\gamma$ decay, $BR(\mu\rightarrow e\gamma) < 5.7\times 10^{-13}$, representing the shaded area in the plot.  The important outcome in this analysis is that there is plenty of space to obtain  neutrino masses at eV scale through the ISSII mechanism with new physics, in the form of Higgs triplets, with mass in the EW scale until TeV scale. Perceive that the smaller  $v_\Delta$ is, the bigger $M_\Delta$ gets. For example, for $v_\Delta $ at sub-eV scale implies $M_\Delta$ above TeV. As consequence,  because the ILC is planed  to run firstly in  500 and 1000 GeV,  then only Higgs triplets with mass at electroweak scale can be probed in the first running of the ILC. However, according to the  bound in FIG.~\ref{fig1}, Higgs triplets with mass at electroweak scale  requires $v_\Delta$ around $10^{-8}$ GeV. This case leads to tiny Yukawa couplings which may difficult  the probe of these scalars at the ILC or LHC. 

We stress  that, in the regime of  energy where the ISSII mechanism is valid, $\delta^{++}$ and $H^+$ are practically  degenerated in mass, see  Eqs.~(\ref{hpmass}) and (\ref{hppmass}). Consequently,  $\delta^{++}$ may decay into a pair of charged leptons and/or charged gauge bosons, $W^\pm$. However, the decay into $W^\pm$ is strongly suppressed in our model due to the smallness of $v_\Delta$.  This is so because the coupling of $W^{\pm}$ with the charged scalars $\delta^{++}$  is proportional to $v_\Delta$.Thus, for example, for $v_\Delta=10$eV we have $BR(\delta^{--}\rightarrow W^- W^- )\approx 2\times 10^{-9}$.  Moreover, in what concerns the branching ratio of $\delta^{++}$ into pair of leptons, our calculations considered two situations. In the case of NH, $\delta^{++}$ will  decay preferentially into a  $\mu^+\tau^+$  pair, with BR around 46\% and into pairs of  $\mu^+ \mu^+$  and $\tau^+ \tau^+$, with   BR$\approx 23\%$  each.  For the case of IH, $\delta^{++}$ will decay preferentially into a  $e^+ e^+$ pair, with  $BR\approx 46\%$, while  it decays 27\% of the times into the  $e^+\tau^+$ pair and 15\% into the $\mu^+ \mu^+ $ pair. After all this analysis we are read to explore the signature of the ISSII at the ILC and LHC. We do this in the next section. 
\begin{figure}[tb]
\begin{center}
\includegraphics[scale=0.35]{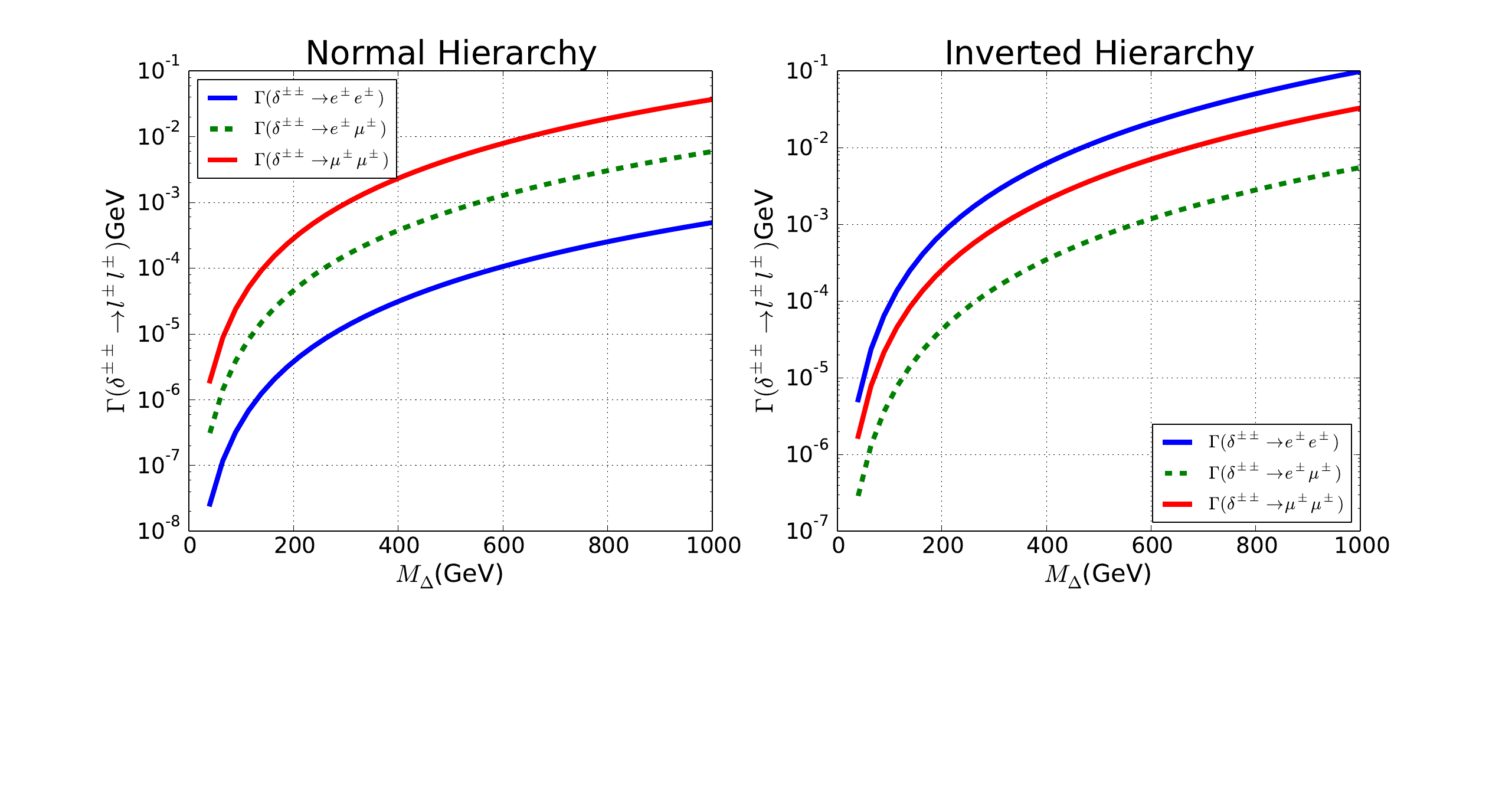}
\caption{ $\Gamma(\delta^{\pm \pm}\rightarrow l^{\pm \pm}l^{\pm \pm})$ vs $M_\Delta$  for the cases of  NH and IH with pair of charged leptons as final product }
\label{fig2}
\end{center}
\end{figure}
\section{Signature of the ISSII mechanism at the ILC and LHC}
\label{sec6}

First of all, it is necessary to say that the work done in this section  is complementary to all other works done focusing at the LHC search for the $\delta^{ ++}$ LHC\cite{spectrumDelta} where a direct search  in the  CMS and ATLAS colliders   has been already  performed  and the result was the   limit  $m_{\delta^{++}}> 459$~GeV for $\delta^{++}$ decaying 100\% into $\mu^+ \mu^+$ pairs~\cite{deltaLHC}.  Investigation of direct detection of $\delta^{++}$ at ILC through  diboson decay scenario ($W^+W^+$) has also been  considered in \cite{ILC}.  As we have discussed in the end of the last section, in the regime of validity of the ISSII mechanism we are adopting here,   the detection of $\delta^{++}$ must occur through dilepton decay scenarios, $e^+e^- \rightarrow \delta^{++} +\delta^{--} \rightarrow l^+ l^+ l^- l^-$, where $l=e,\,\mu$ with BR of the same order of magnitude for the most important channels, but surely not 100\% into $\mu^+ \mu^-$ pairs. Recalling that diboson decays are suppressed in the regime of validity of the  ISSII mechanism ( which means to take $v_\Delta$ around eV scale)  then the LHC  and ILC investigations done in \cite{deltaLHC} cannot be applied for the case of ISSII mechanism.  This justify the search for $\delta^{++}$ at the LHC and ILC in the regime of energy of the ISSII mechanism.  

The necessary ingredients needed to develop our proposal are the Yukawa interactions among $\delta^{++}$  and the leptons, given in Eqs. (\ref{Yukawacoupling}), whose couplings is given in  (\ref{yukawarelations}), and the interactions of $\delta^{++}$ with the standard gauge bosons, whose couplings are given  in the TABLE. \ref{table1}. Other interactions we use  involve the standard particles and  may be found in any textbooks and reviews.  
\begin{table}[!h]
\begin{tabular}{clcl}
\hline 
INTERACTION & COUPLING  \\ 
\hline 
$\delta^{++} W^+_\mu W^+_\nu$ & $-i \sqrt{2} g^2 v_\Delta g_{\mu \nu}$  \\ 
\hline 
 $\delta^{++} \delta^{--}\gamma_\mu$ & $-2 i e (P_{\delta^{++}} -P_{\delta^{--}})_\mu$ \\ 
\hline 
 $\delta^{++} \delta^{--}Z^0_\mu$ & $-2 i e \cot(2\theta_W) (P_{\delta^{++}} -P_{\delta^{--}})_\mu$ \\
\hline
\end{tabular} 
\caption{Interactions and couplings of $\delta^{++}$ with the standard gauge bosons. }
\label{table1}
\end{table}

Firstly we present the results for the ILC with $\sqrt{s}= 500$ GeV and $1$ TeV. Next, we present the results for the  run II of the LHC. Our results take into account the  cases of normal and inverted hierarchies. Our numerical calculations are based in the following routine: the model was adapted in FeynRules \cite{FRules}. The UFO output generated by FeynRules was
imported in MADGRAPH5 \cite{MG1,MG2,MG3} to produce events for each channel. The samples were based on 150 thousand events generated with the values of $v_{\Delta}$, $M_\Delta$ and and $\mu$ described in the tables. \ref{table2} and \ref{table3}. The LHE files were passed through PYTHIA6 \cite{pythia} for showering and hadronization. Jets were reconstructed with FastJet \cite{fastjet} using an anti-k algorithm with a cone size R = 0.4. The resulting hadronized events were analyzed using MadAnalysis 5 \cite{madanalysis1,madanalysis2}. In MadAnalysis the sample generated with $\sqrt{s} = 500$ GeV are loaded and set with $500$fb$^ {-1}$ while  $\sqrt{s} = 1$ TeV are loaded and set with $1000$fb$^ {-1}$ \cite{ILCwhitpaper}.
For LHC the sample generated with $\sqrt{s} = 13$TeV are loaded and set with $41.07$fb$^ {-1}$ luminosity delivered for oct 2016\cite{luminosity}. The following cuts were applied: $P^j_T > 20$ GeV,
$ P_ tT^l > 10$ GeV,$ P_T^a > 10$ GeV, $| \eta_j |<4.5$, $| \eta_l |< 2.5$, $ÆR_{ll} > 0.4$ and $ÆR_{jj} > 0.4$.

\subsection{ILC}
\begin{figure}[tb]
\begin{center}
	\includegraphics[scale=0.40]{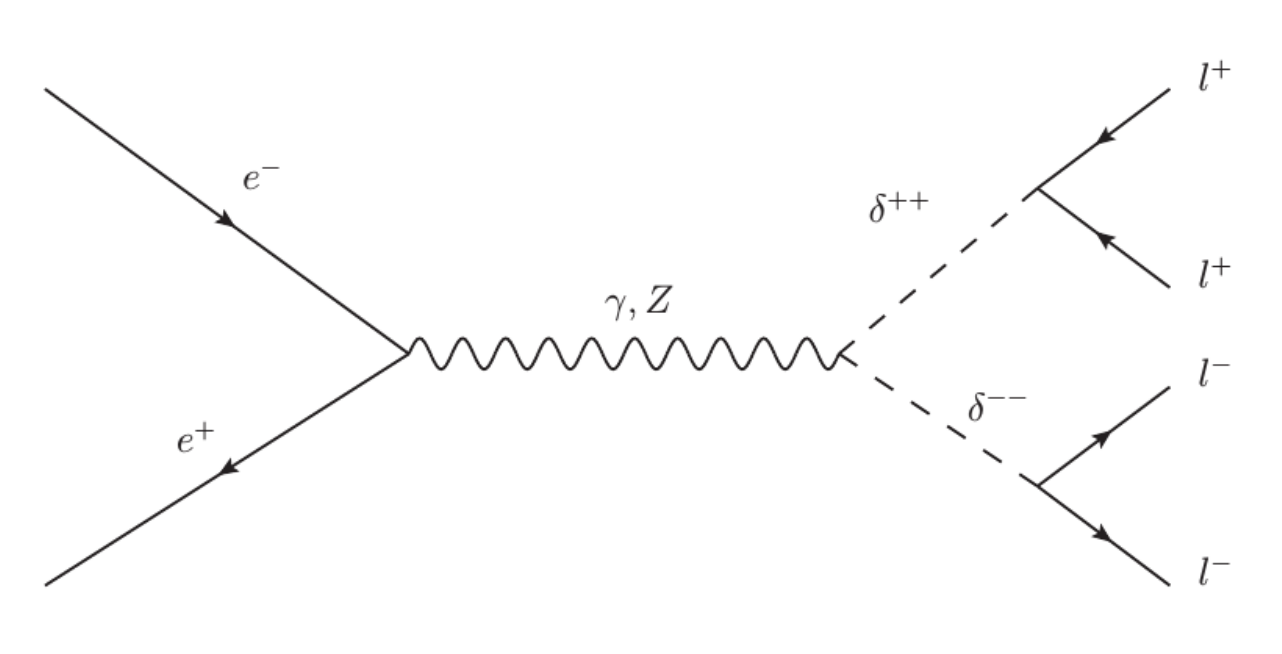} 
\caption{Dominant contributions    for the  processes $e^+ + e^- \rightarrow \delta^{++} +\delta^{--} \rightarrow l^+ l^- l^+ l^-$.}
 \label{figILC}
\end{center}
\end{figure}

At the ILC, we consider the processes $ e^+ + e^- \rightarrow \delta^{++} +\delta^{--} \rightarrow l^+ l^- l^+ l^-$ with  $l=e, \mu$ as final product. The dominant contributions  are displayed in FIG. (\ref{figILC}).  Remembering that we took $m_1=0$  in the  NH case  and $m_3=0$ in the IH one. Once $m_{1,2,3}$ and  $\theta_{12}$, $\theta_{23}$ and $\theta_{13}$ are fixed, the Yukawa couplings in Eq. (\ref{yukawarelations}) get  depending exclusively  on $v_\Delta$, which, in turn, is related to the values of $\mu$ and $M_\Delta$. In the TABLE. II  we show the set of values of these parameters that we made use in our numerical calculations for the analysis at the ILC.
\begin{table}[!h]
\begin{tabular}{|c|c|c|c}
\hline
$M_\Delta$(GeV) & $\mu (GeV)$ & $v_{\Delta}$(GeV) \\
\hline 
100 & $2.8676699999999999*10^{-9}$ & $1.2293140255665008*10^{-8}$ \\ 
\hline 
115 & $3.3149700000000002*10^{-9}$ & $1.0745276253585174*10^{-8}$ \\ 
\hline 
130 & $3.7123300000000001*10^{-9}$ & $9.4165872339574946*10^{-9}$ \\ 
\hline 
145 & $4.1959100000000002*10^{-9}$ & $8.5550750426766403*10^{-9}$ \\ 
\hline 
160 & $4.5535700000000000*10^{-9}$ & $7.6251030172672466*10^{-9}$ \\ 
\hline 
175 & $5.0224300000000000*10^{-9}$ & $7.0302611643695496*10^{-9}$ \\ 
\hline 
190 & $5.4197200000000000*10^{-9}$ & $6.4358114458721346*10^{-9}$ \\ 
\hline 
200 & $5.7207500000000001*10^{-9}$ & $6.1309340089336988*10^{-9}$ \\ 
\hline
205 & $5.8452300000000001*10^{-9}$ & $5.9624883860371309*10^{-9}$ \\ 
\hline 
230 & $6.5816599999999998*10^{-9}$ & $5.3335139463683973*10^{-9}$ \\ 
\hline 
260 & $7.3835499999999998*10^{-9}$ & $4.6822240123646648*10^{-9}$ \\ 
\hline 
290 & $8.4257699999999999*10^{-9}$ & $4.2948427541542566*10^{-9}$ \\ 
\hline 
320 & $9.1908499999999995*10^{-9}$ & $3.8475952970005223*10^{-9}$ \\ 
\hline 
350 & $9.9895800000000006*10^{-9}$ & $3.4957857213720639*10^{-9}$ \\ 
\hline 
380 & $1.0810200000000000*10^{-8}$ & $3.2092252409795597*10^{-9}$ \\ 
\hline 
410 & $1.1612700000000000*10^{-8}$ & $2.9614142164009535*10^{-9}$ \\
\hline
\end{tabular} 
\caption{Values of $M_\Delta$, $\mu$ and $v_\Delta$ allowed by the Eq. (\ref{ISSequation}) that we used in our analysis at the ILC.}
\label{table2}
\end{table}

\begin{figure}[tb]
\begin{center}
	\includegraphics[scale=0.17]{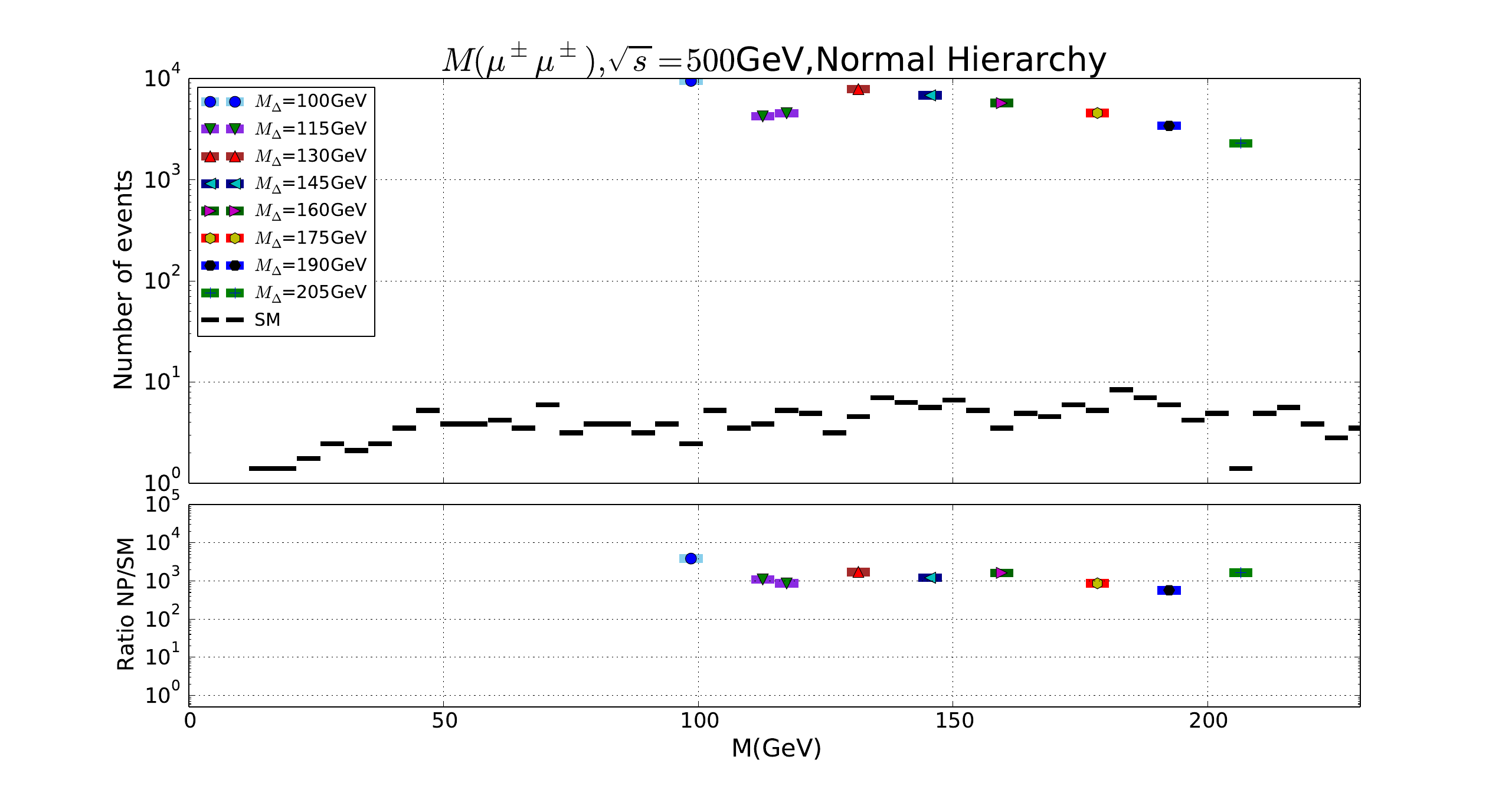} \quad
	\includegraphics[scale=0.17]{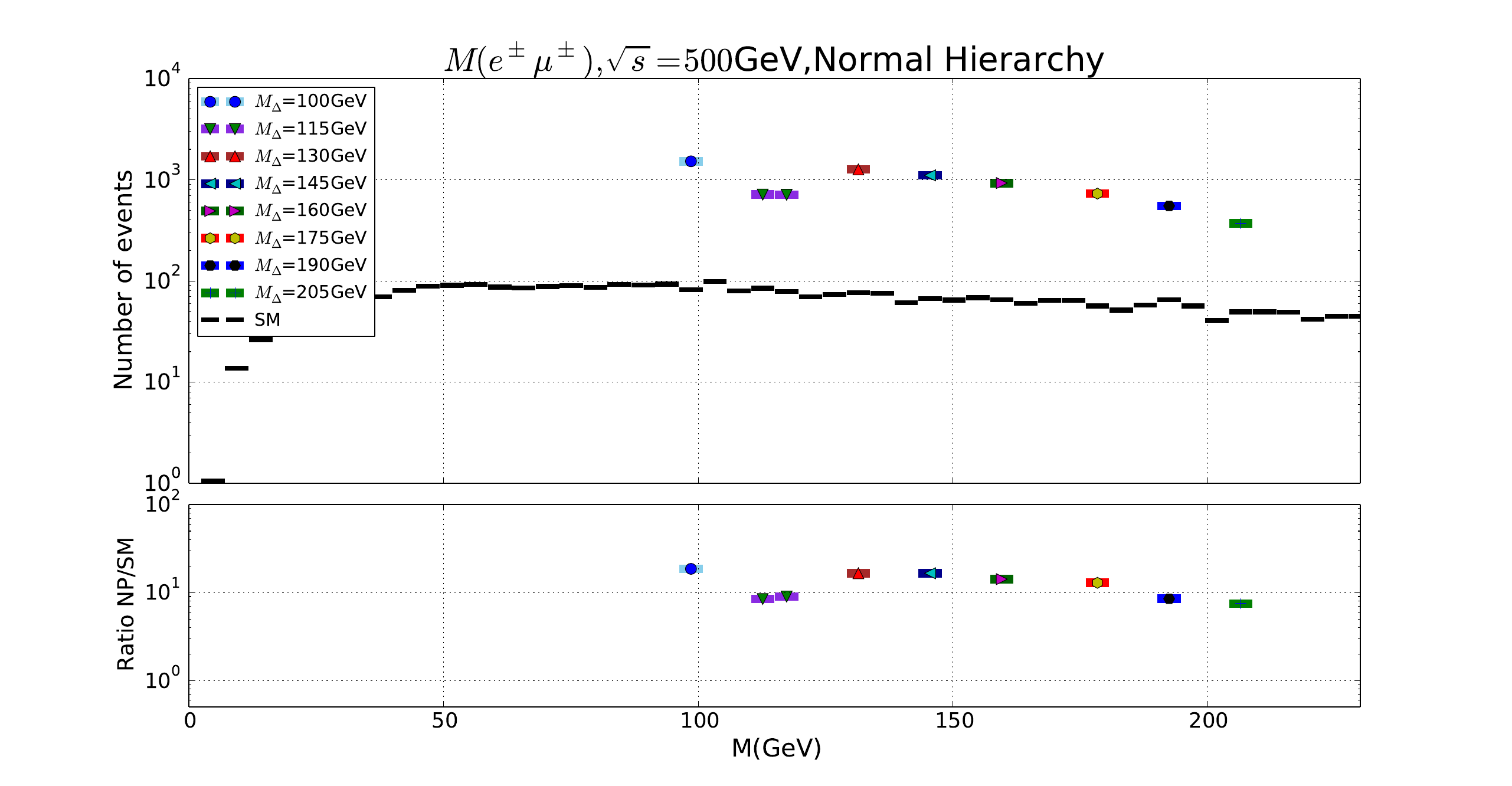} \quad
	\includegraphics[scale=0.20]{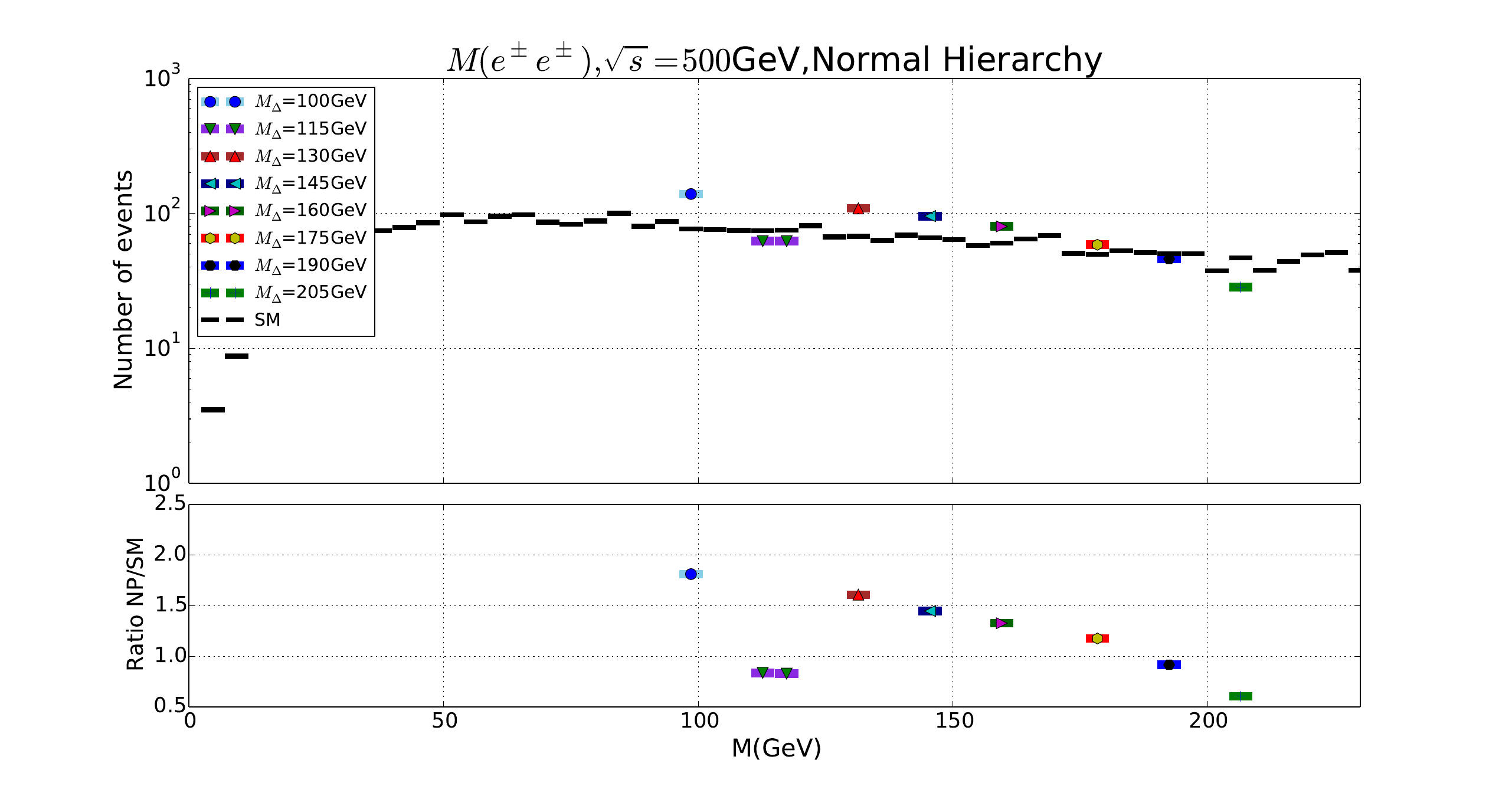}
\caption{The  process $e^+ + e^- \rightarrow \delta^{++} +\delta^{--} \rightarrow l^+ l^- l^+ l^-$ (with $l=e, \mu$) at the ILC for the NH case with $\sqrt{s}=500$ GeV.}
 \label{fig3}
\end{center}
\end{figure}

\begin{figure}[tb]
\begin{center}
	\includegraphics[scale=0.17]{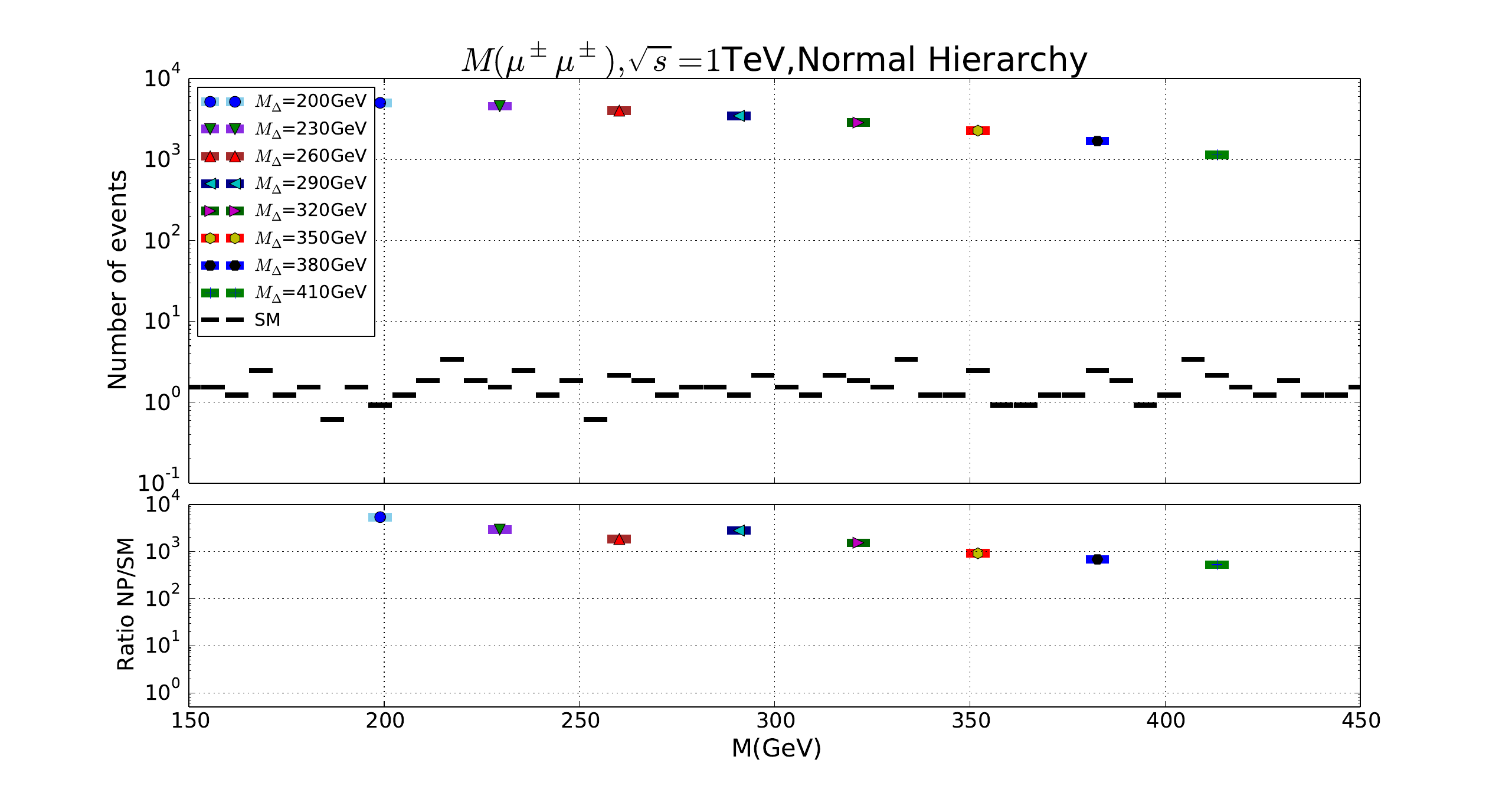} \quad
	\includegraphics[scale=0.17]{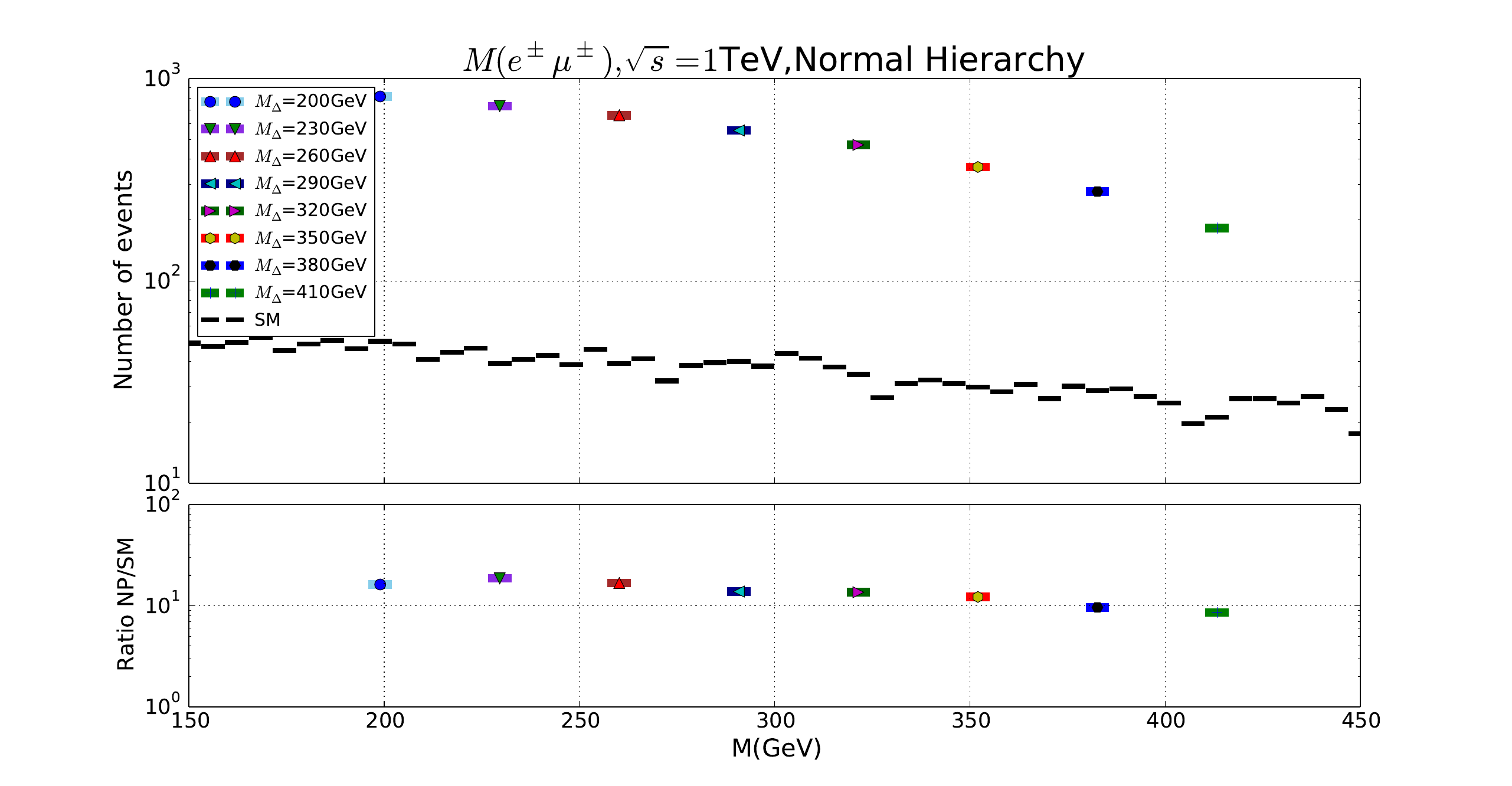} \quad
	\includegraphics[scale=0.20]{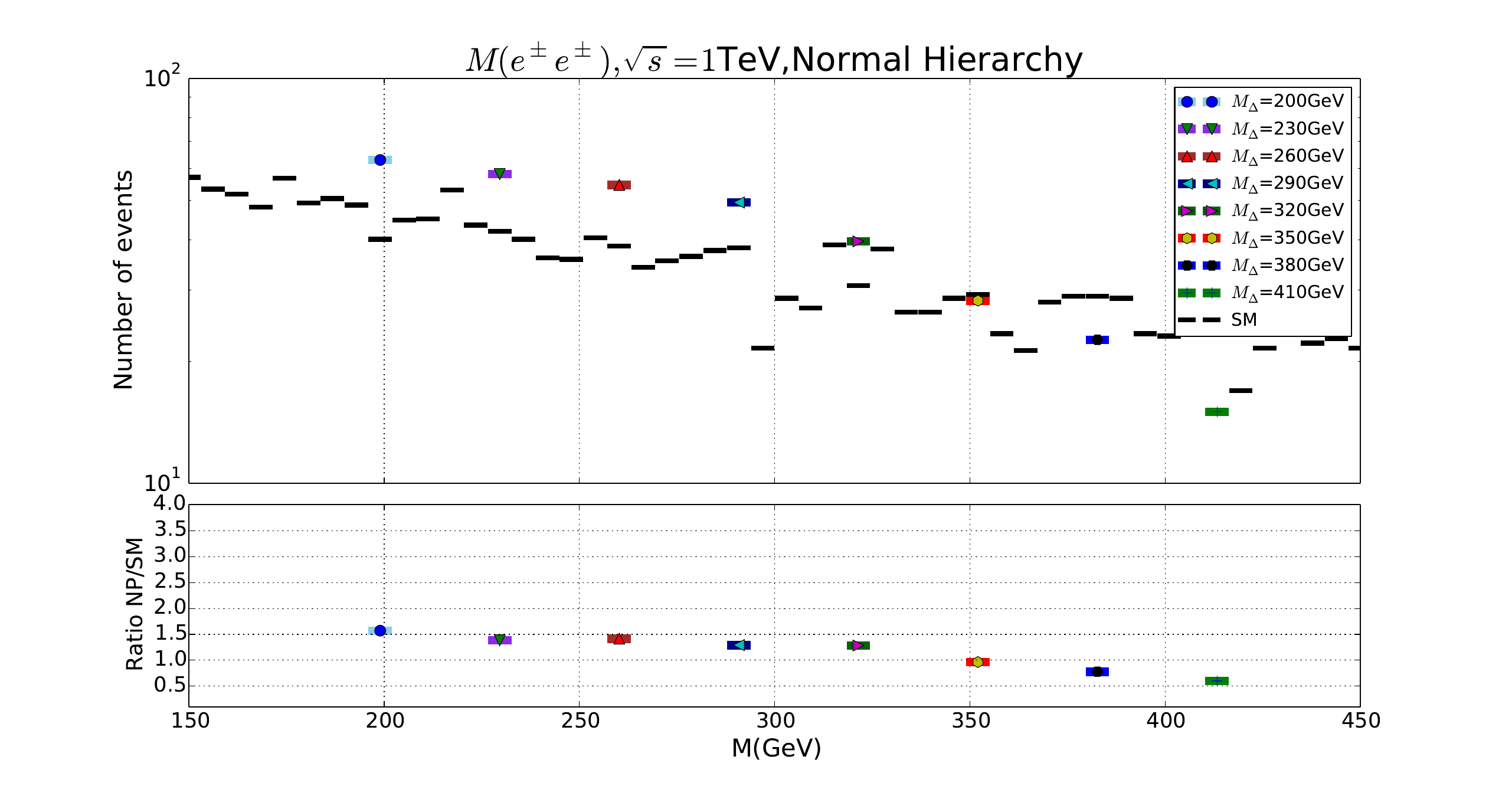}
\caption{The  process $e^+ + e^- \rightarrow \delta^{++} +\delta^{--} \rightarrow l^+ l^- l^+ l^-$  (with $l=e, \mu$) at the ILC for the NH case with  $\sqrt{s}=1$ TeV.}
 \label{fig4}
\end{center}
\end{figure}

\begin{figure}[tb]
\begin{center}
	\includegraphics[scale=0.17]{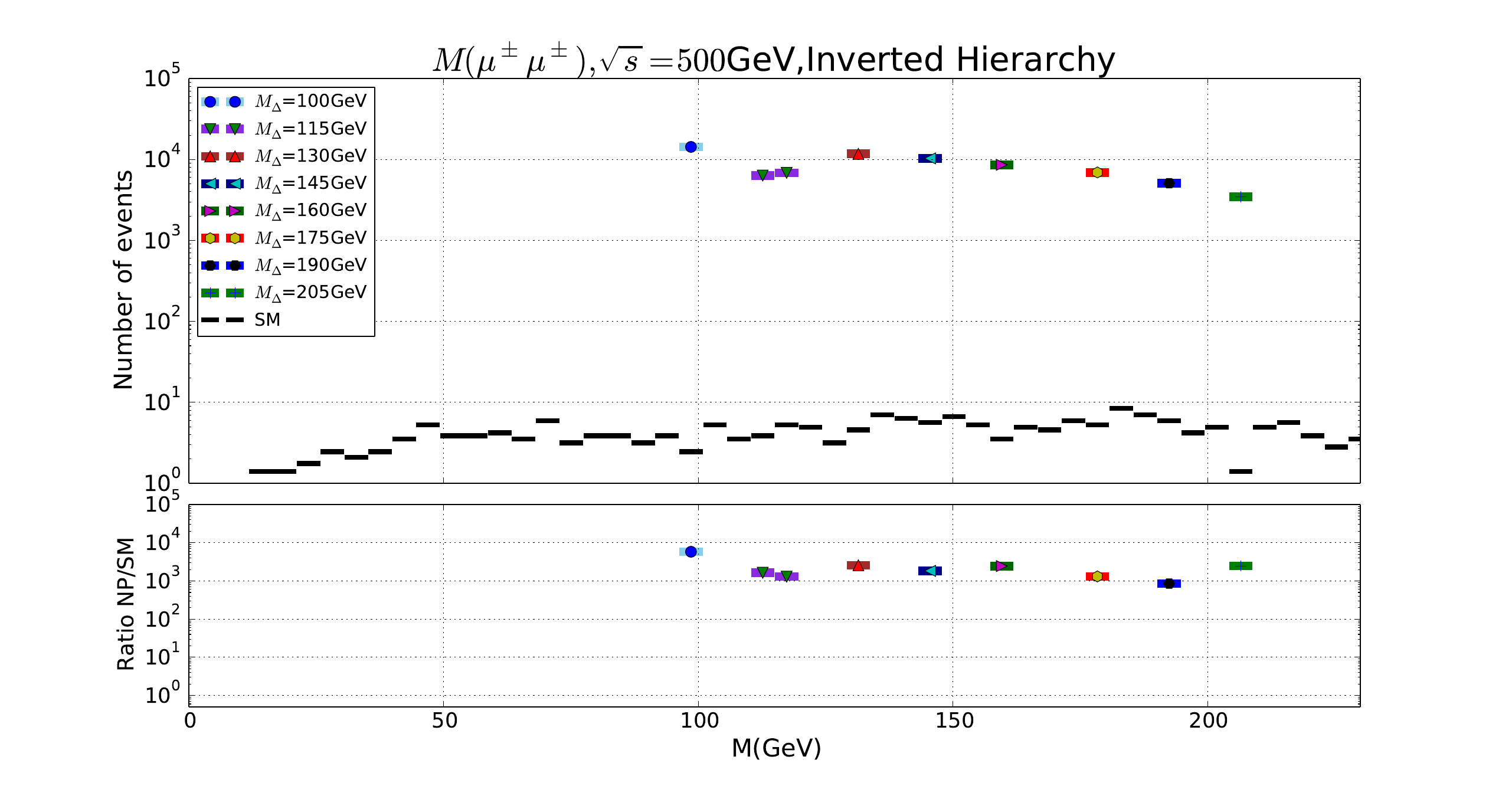} \quad
	\includegraphics[scale=0.17]{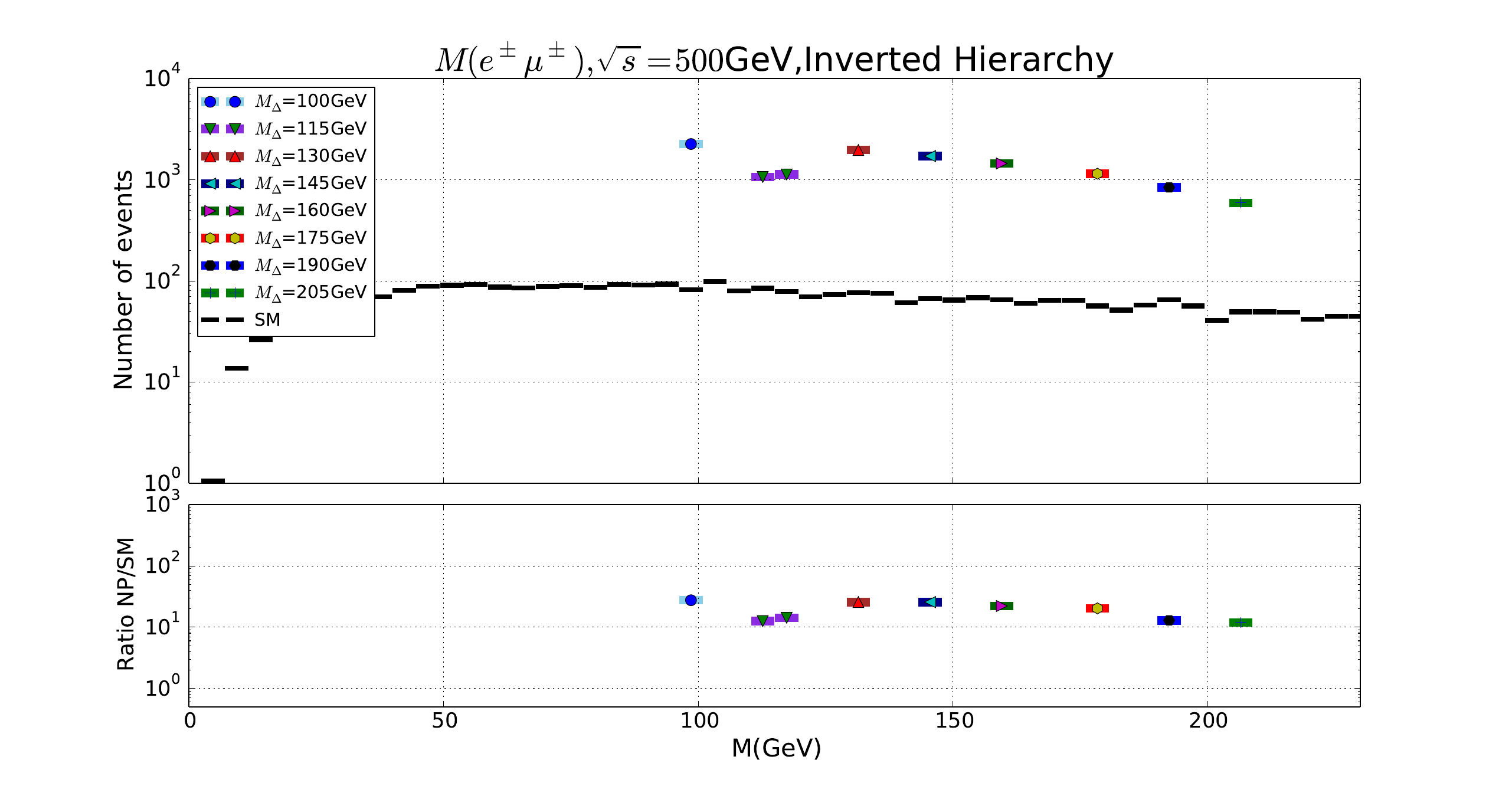} \quad
	\includegraphics[scale=0.20]{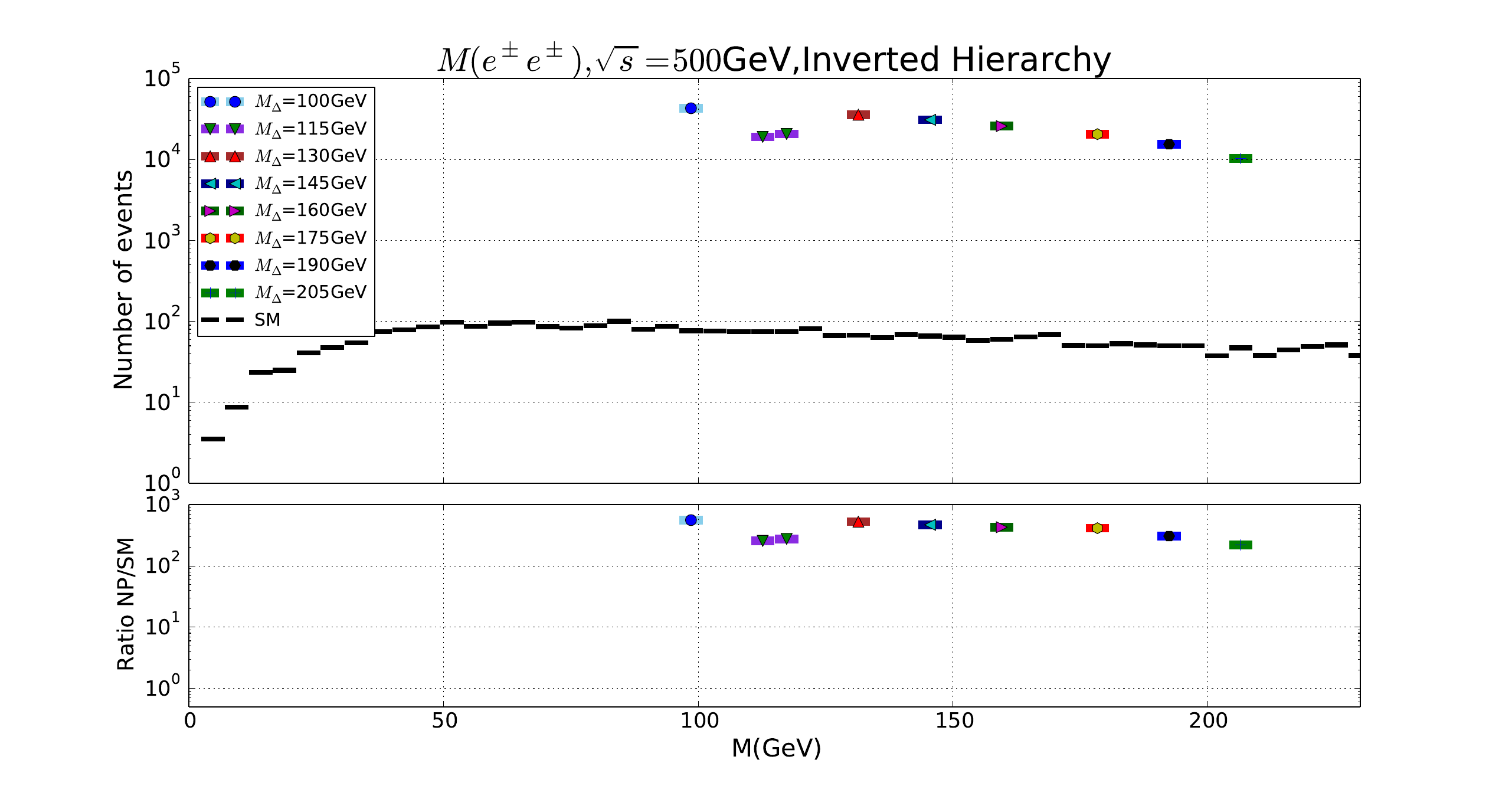}
\caption{The  processes $e^+ + e^- \rightarrow \delta^{++} +\delta^{--} \rightarrow l^+ l^- l^+ l^-$   (with $l=e, \mu$) at the ILC for the IH case with $\sqrt{s}=500$ GeV.}
 \label{fig5}
\end{center}
\end{figure}

\begin{figure}[tb]
\begin{center}
	\includegraphics[scale=0.17]{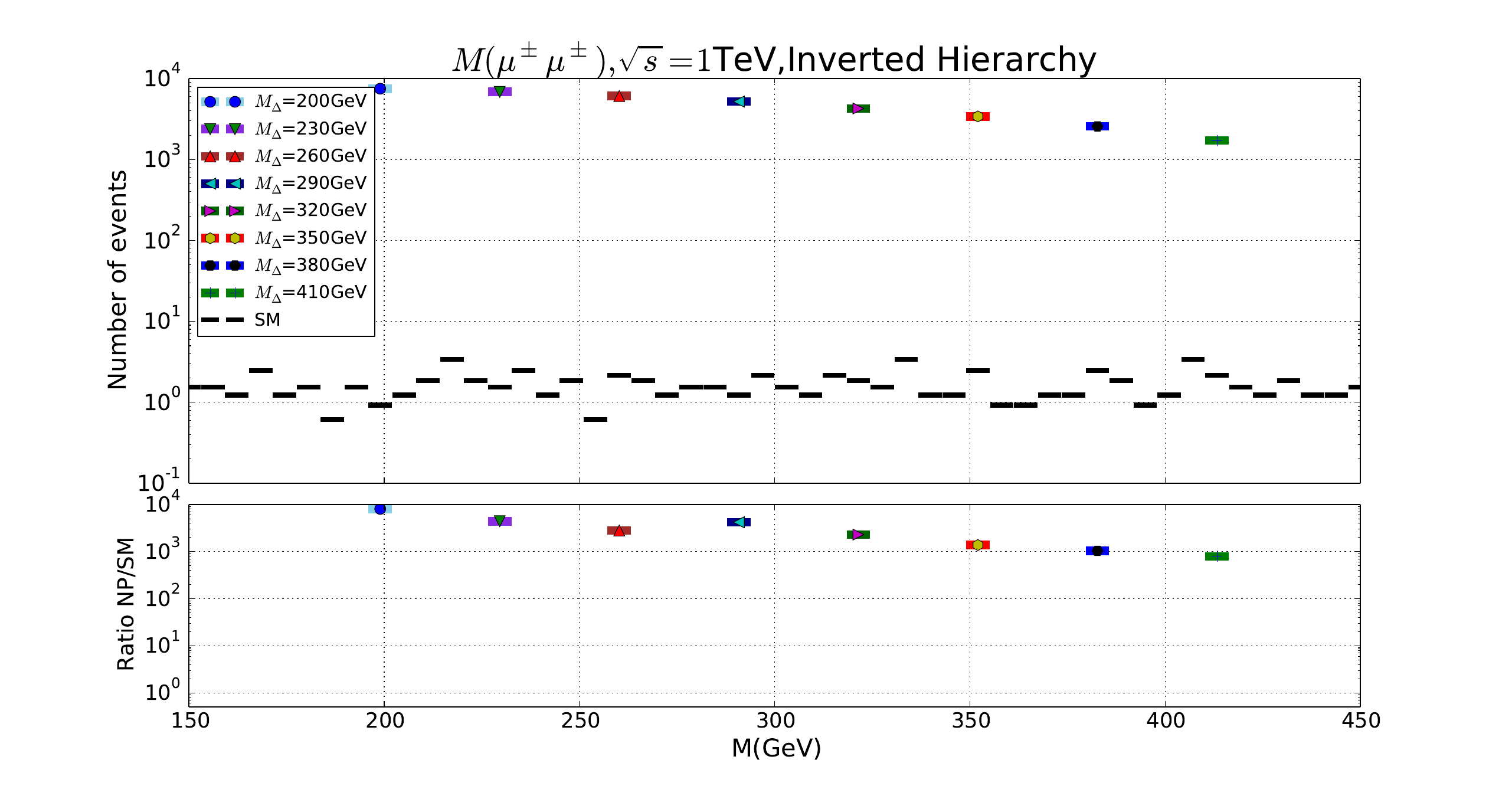} \quad
	\includegraphics[scale=0.17]{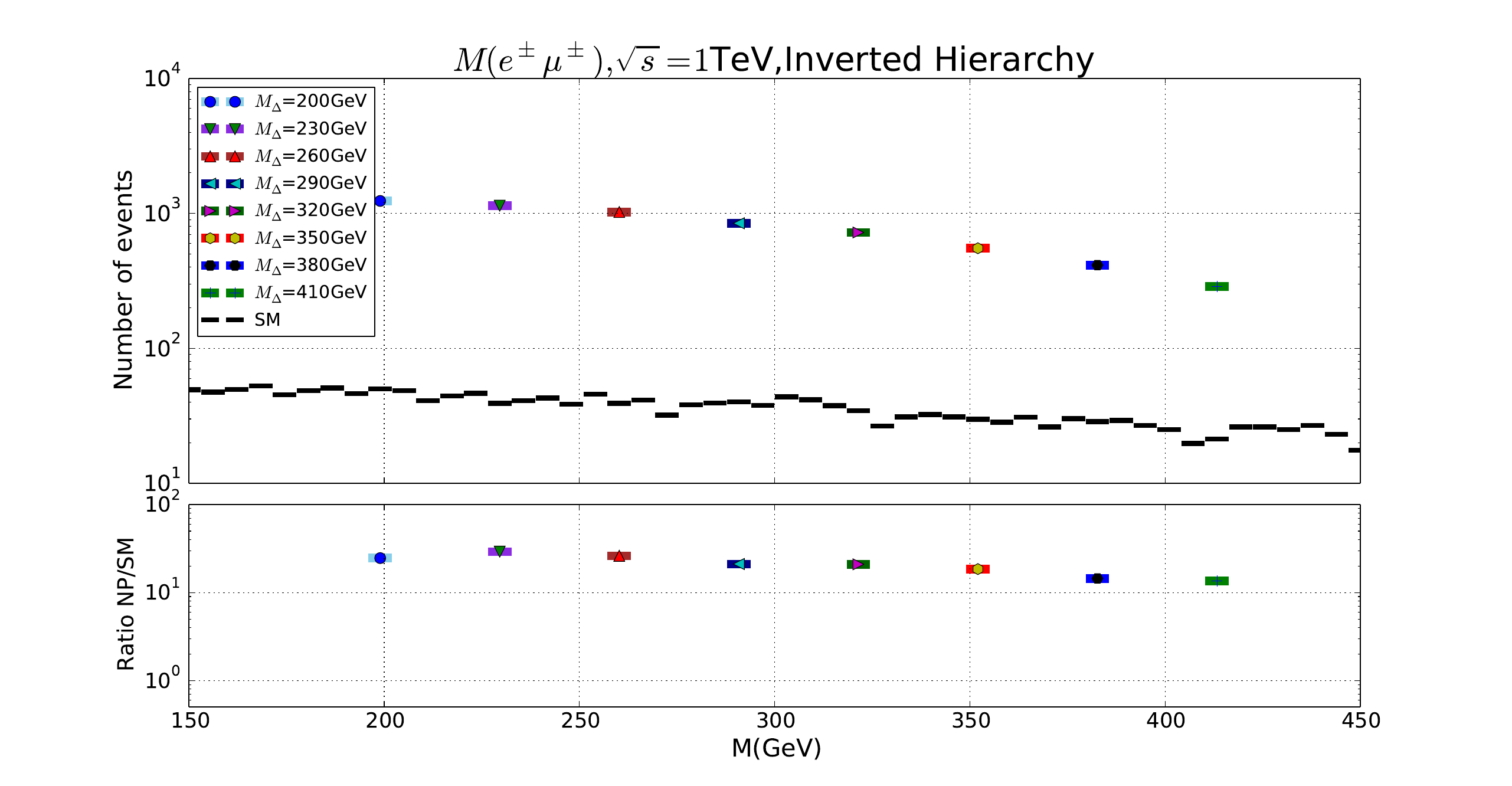} \quad
	\includegraphics[scale=0.20]{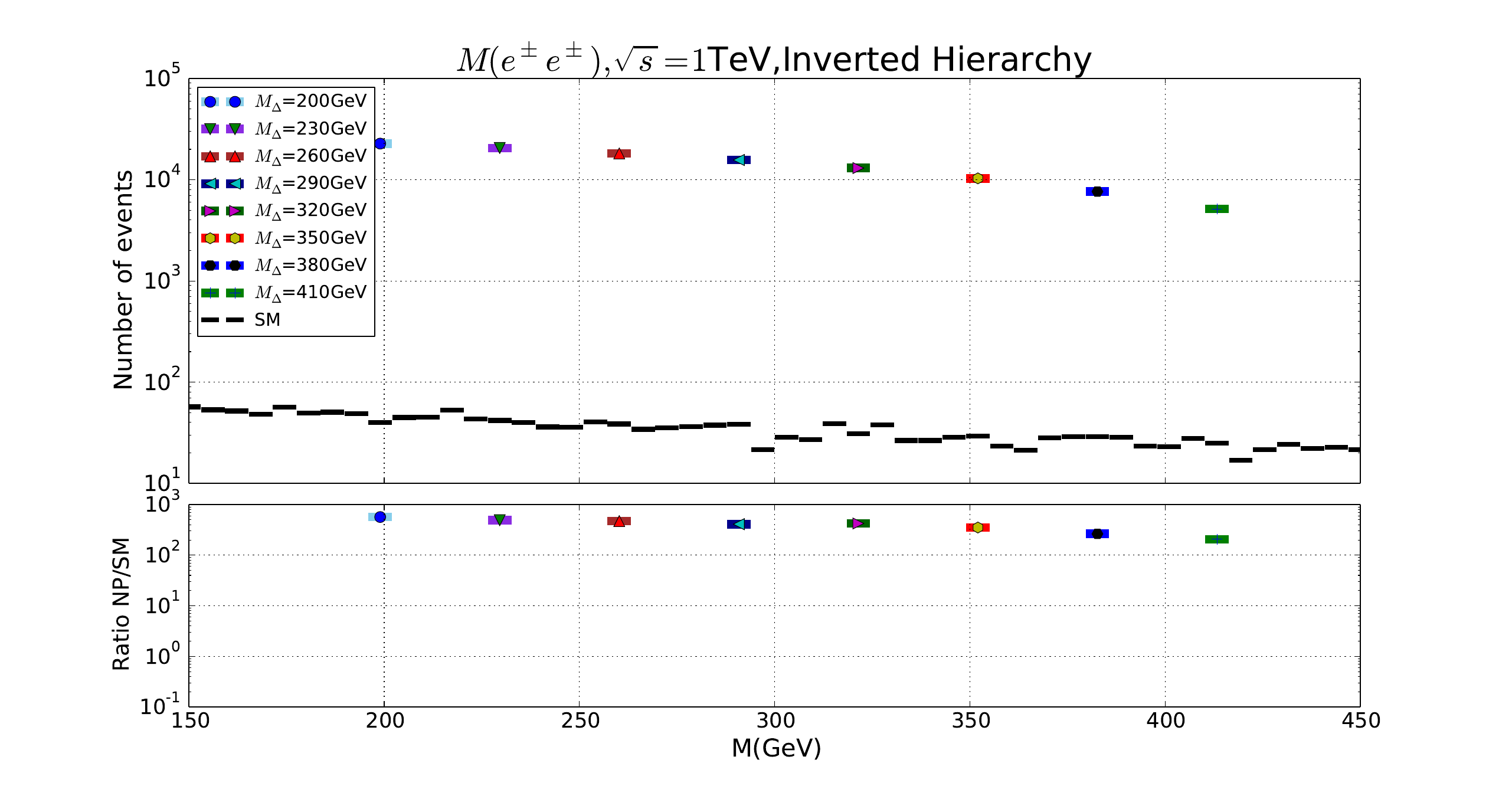}
\caption{The  process $e^+ + e^- \rightarrow \delta^{++} +\delta^{--} \rightarrow l^+ l^- l^+ l^-$  (with $l=e, \mu$) at the ILC for the  NH case with $\sqrt{s}=1$TeV.}
 \label{fig6}
\end{center}
\end{figure}

In  FIG. (\ref{fig3}) and  FIG. (\ref{fig4}) we display our  results for the NH case, while in  FIG. (\ref{fig5}) and  FIG. (\ref{fig6}) we do the same for the IH one. In both cases  the background is due to the processes $ e^+ + e^- \rightarrow Z^0/\gamma + Z^0/\gamma \rightarrow l^+ l^- l^+ l^-$. In each plot we present the number of events as function of the invariant mass of pairs of charged leptons and, in the bottom part, we present  the ratio among the number of events,  due to exclusively the  new physics (NP), and the background ( the  number of events due to  the SM).  For the case of 500 GeV, we considered $M_{\Delta}$ varying from 100 until 205 GeV, while for the case of 1TeV, we considered $M_{\Delta}$ varying from 200 until 410 GeV. 

In the NH case, displayed in FIG. (\ref{fig3}) and  FIG. (\ref{fig4}), we see that   the process $e^+ + e^- \rightarrow \delta^{++} +\delta^{--} \rightarrow \mu^+ \mu^+ \mu^- \mu^-$ is the dominant  one. This is expected according to the profile of the decay width of $\delta^{++}$, as showed in FIG. (2).  The ratio NP/SM for   $\mu^{\pm }\mu^{\pm }$ as final product  is around $10^3$ which means that the ILC will be very efficient in producing  such processes. On the other hand, the process $e^+ + e^- \rightarrow \delta^{++} +\delta^{--} \rightarrow e^+ e^+ e^- e^-$ is produced in  the same order that of the background. This is also very clear in the ratio NP/SM. The process  $e^+ + e^- \rightarrow \delta^{++} +\delta^{--} \rightarrow e^+ \mu^+ e^- \mu^-$ also  presents a good production once  the number of events  is 10 times higher than the number of events provided by the background. This analysis is applied to both   cases of 500 GeV and 1TeV. 

On the contrary,  in the IH case,  which are displayed in FIG. (\ref{fig5}) and  FIG. (\ref{fig6}) , the dominant  process  is now  $e^+ + e^- \rightarrow \delta^{++} +\delta^{--} \rightarrow e^+ e^+ e^- e^-$ with the number of events reaching the order of $10^5$. However, this case provides the biggest background which  achieve the order of magnitude of  $10^2$. This explains why the ratio NP/SM is  smaller than the previous case in FIGs. (\ref{fig3}) and (\ref{fig4}) .  We also stress that, in this case, the process $e^+ + e^- \rightarrow \delta^{++} +\delta^{--} \rightarrow \mu^+ \mu^+ \mu^- \mu^-$ is not negligible. Besides the number of event provided by this process is smaller than the $e^+ + e^- \rightarrow \delta^{++} +\delta^{--} \rightarrow e^+ e^+ e^- e^-$ case, however its background is one order of magnitude smaller than the dominant case. 

Thus we conclude that   the ILC is  very efficient in probing the signature of the ISSII mechanism discussed here, however we do not expect that it may help in distinguishing the NH case from the IH one. This can be seen clearly  in the process $e^+ + e^- \rightarrow \delta^{++} +\delta^{--} \rightarrow e^+ \mu^+ e^- \mu^-$ for both NH and IH cases. Perceive that the number of events and the ratio NP/SM are very similar in NH and IH cases.


\subsection{LHC}
At the LHC we analyze  the process $pp  \rightarrow \delta^{++} +\delta^{--} \rightarrow l^+ l^- l^+ l^-$ mediated by the gauge bosons $Z^0$ and the photon $\gamma$ for $\sqrt{s}=13$ TeV and luminosity of $41,07$fb$^{-1}$  for the cases of normal and inverted hierarchies.  
\begin{figure}[tb]
\begin{center}
	\includegraphics[scale=0.40]{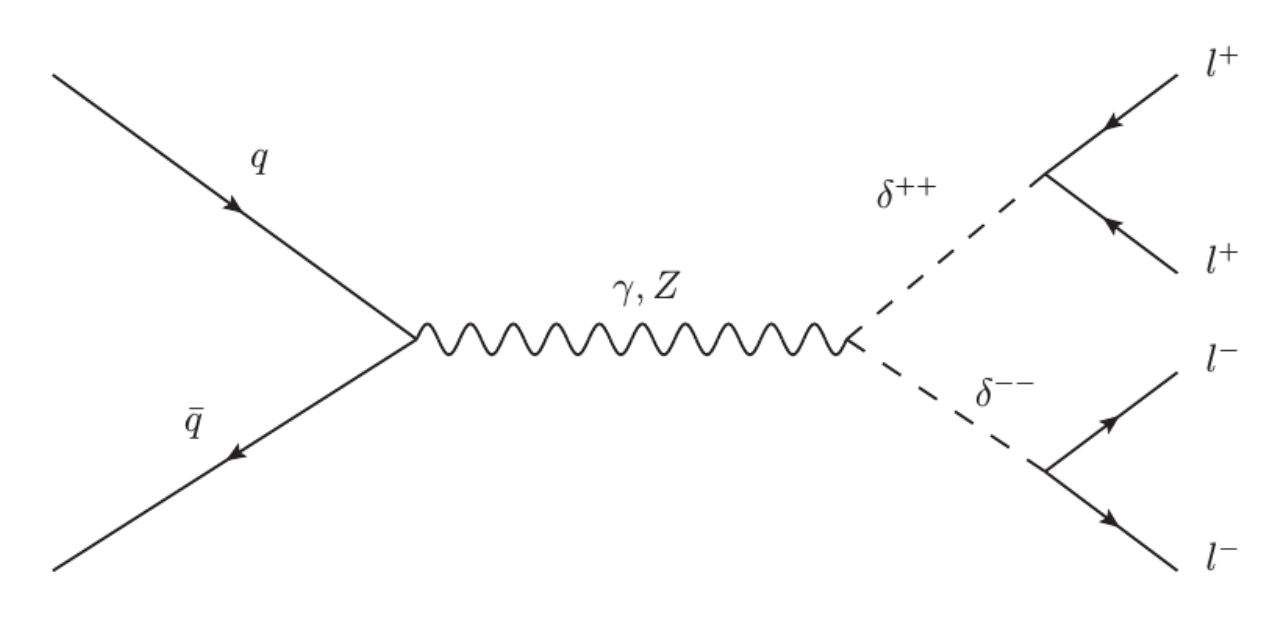} 
\caption{Dominant contributions    for the  processes $q + \bar q \rightarrow \delta^{++} +\delta^{--} \rightarrow l^+ l^- l^+ l^-$.}
 \label{figLHC}
\end{center}
\end{figure}

\begin{figure}[tb]
\begin{center}
	\includegraphics[scale=0.17]{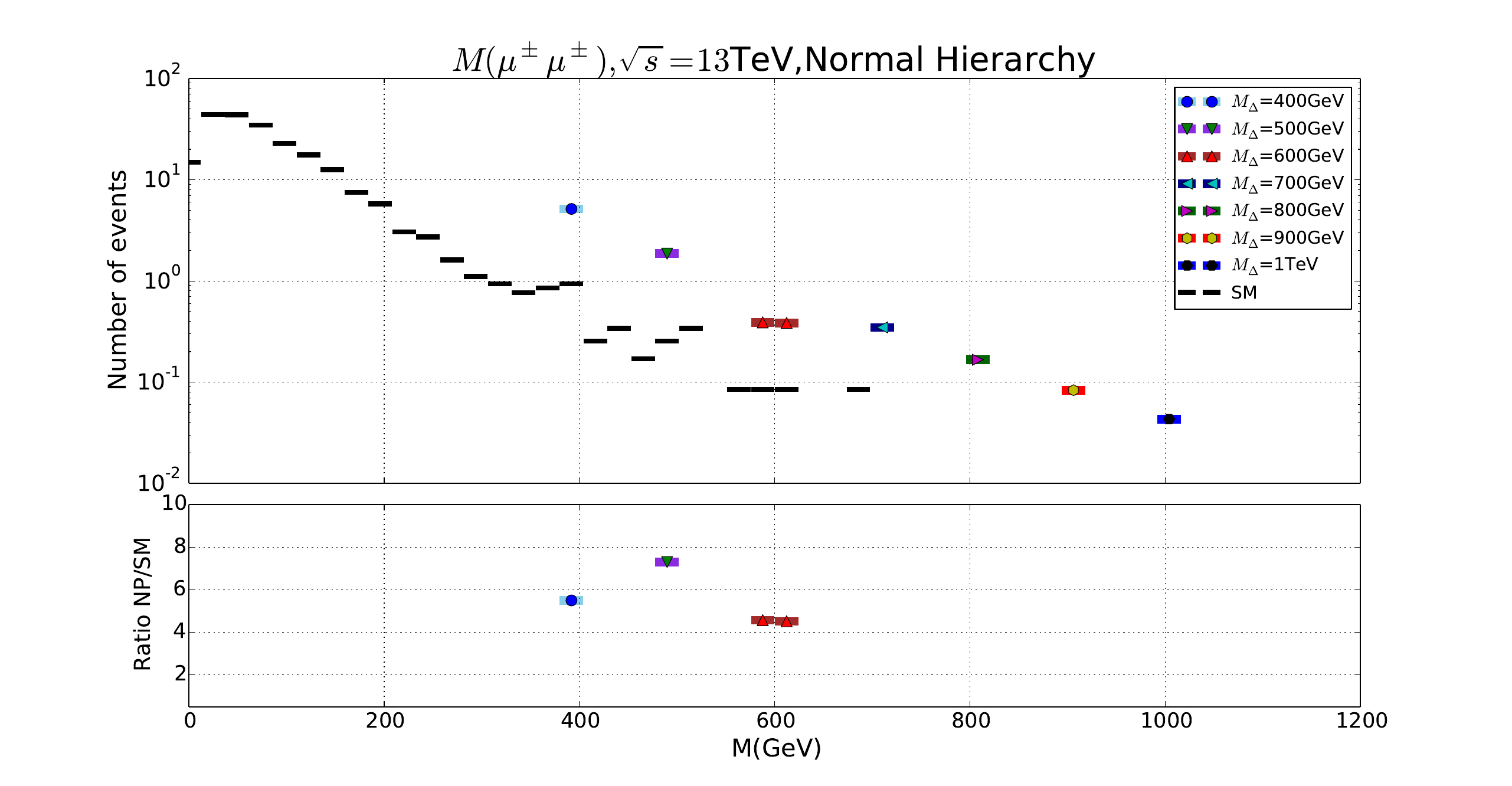} \quad
	\includegraphics[scale=0.17]{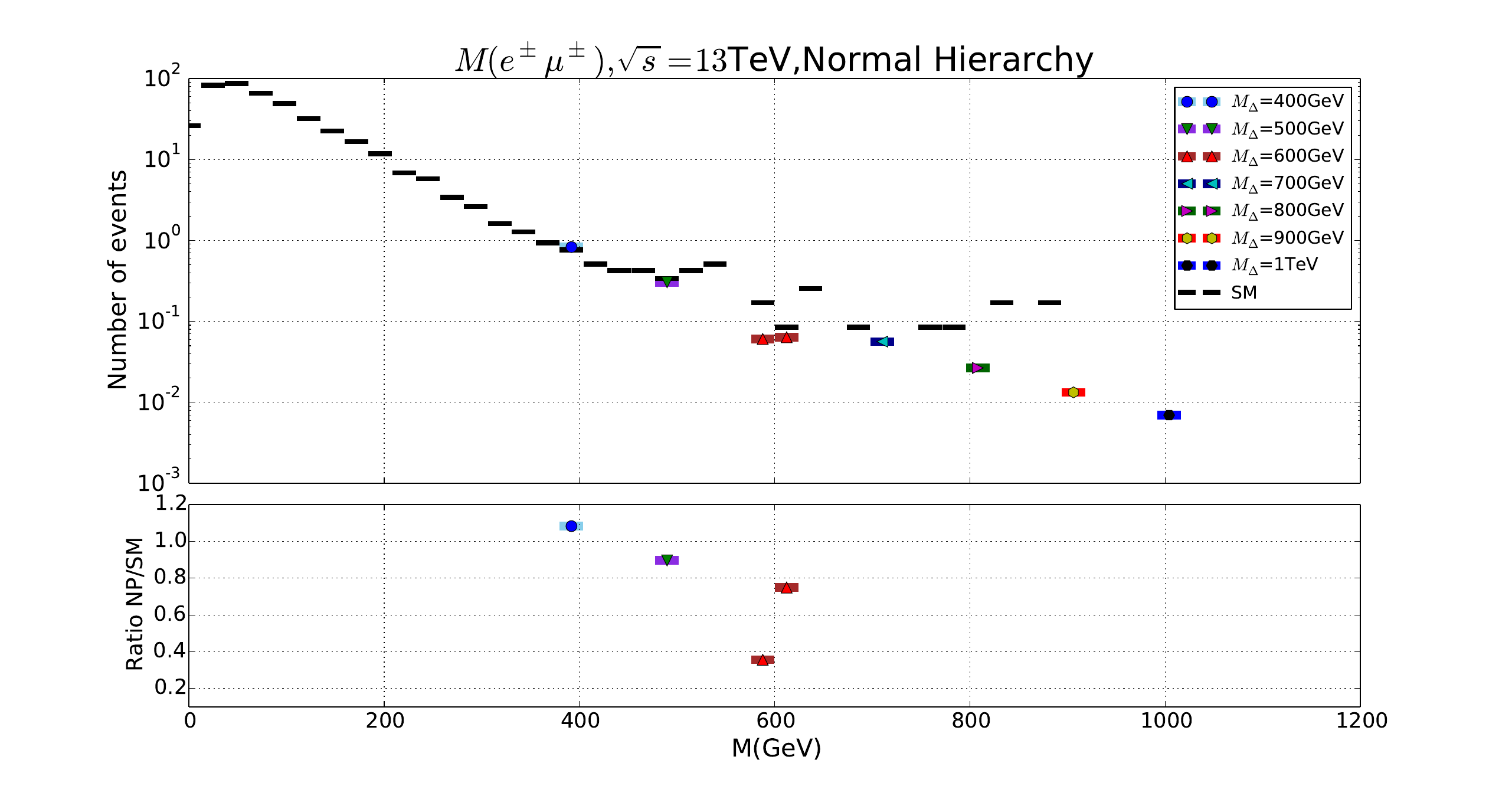} \quad
	\includegraphics[scale=0.20]{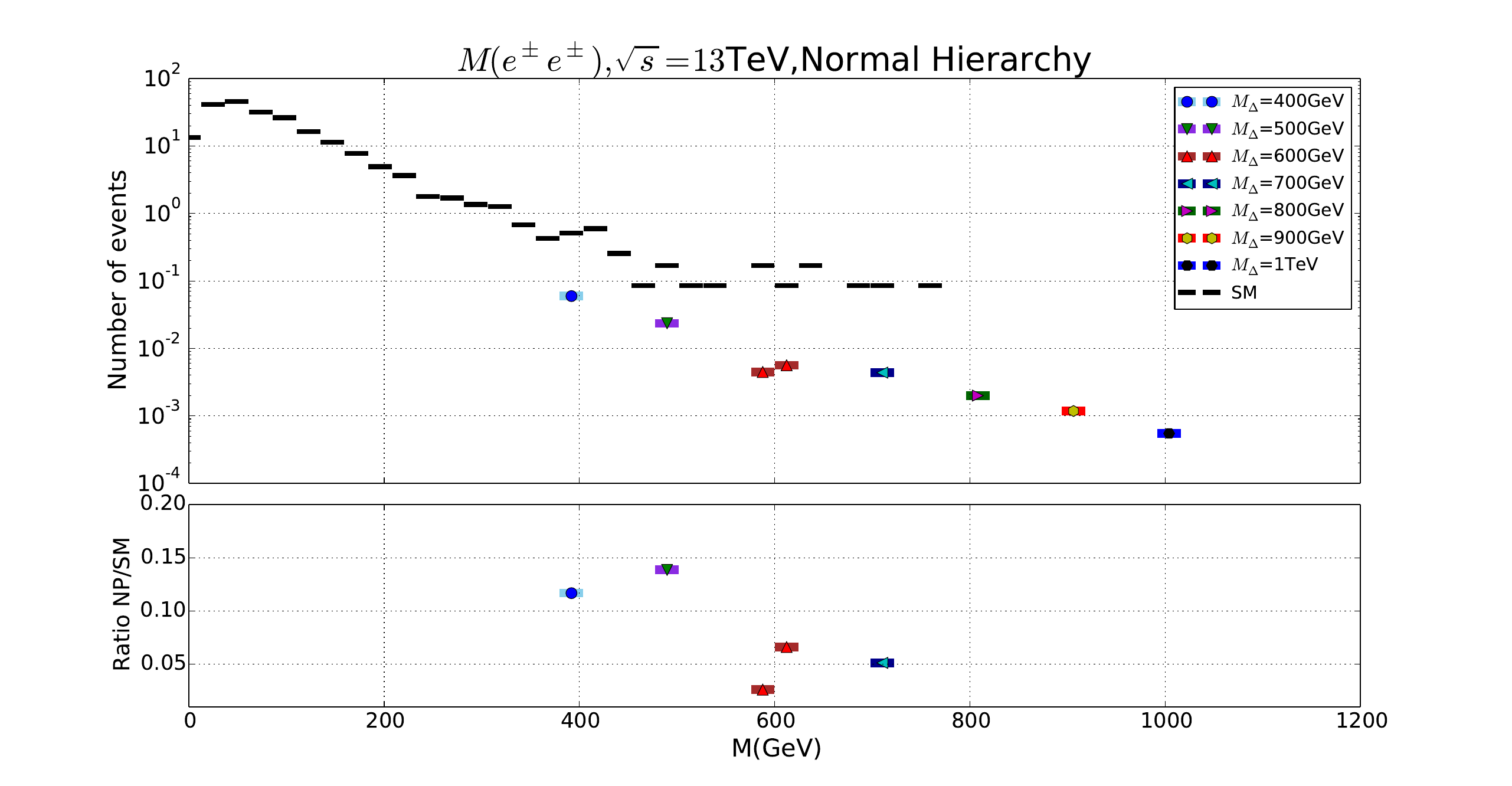}
\caption{The  process $q + \bar q \rightarrow \delta^{++} +\delta^{--} \rightarrow l^+ l^- l^+ l^-$  (with $l=e, \mu$) at the LHC  for the case of NH  with  $\sqrt{s}=13$TeV.}
 \label{fig7}
\end{center}
\end{figure}

In FIG. (\ref{fig7}) we present our results for the NH case. Perceive that in this case only the process resulting in pairs of $\mu^{\pm}\mu^{\pm}$ provides signal above the background.  According to the range of  the mass for $\delta^{++}$ we scan here, the LHC may be sensitive for $\delta^{++}$ with mass in the range 400 until 500 GeV. For the other values the signal is smaller than the background. We may conclude that the LHC is not sensitive for the NH case.

Things are a little different for the IH case. The process of production of pairs of electron-positrons, $e^{\pm}e^{\pm}$, presents sizable sensibility with a significant signal compared to the background with the ratio $NP/SM$ reaching $10^2$ for $\delta^{++}$ with mass in the range 400 until 500 GeV. Thus, according to our analysis we may say that, in spite of the fact that the LHC is not so efficient in producing $\delta^{++}$ as the ILC, however, it may be important  regarding the possibility of discriminating the hierarchy of the neutrino masses. Thus we conclude that, regarding the ISSII mechanism, the conjoint analysis of the  LHC and ILC will be determinant in  probing the signature of the ISSII mechanism and determining the hierarchy of the neutrino masses.
\begin{table}[!h]
\begin{tabular}{|c|c|c|}
\hline
$M_\Delta$(GeV) & $\mu(GeV)$ & $v_{\Delta}$(GeV) \\ 
\hline 
400 & $1.1345800000000000*10^{-8}$ & $3.0398265558956413*10^{-9}$ \\ 
\hline 
500 & $1.4264000000000000*10^{-8}$ & $2.4458790949698630*10^{-9}$ \\ 
\hline 
600 & $1.7314299999999998*10^{-8}$ & $2.0617504641656491*10^{-9}$ \\ 
\hline 
700 & $1.9816699999999998*10^{-8}$ & $1.7336799170914535*10^{-9}$ \\ 
\hline 
800 & $2.2724700000000000*10^{-8}$ & $1.5221303595771494*10^{-9}$ \\ 
\hline 
900 & $2.5821399999999999*10^{-8}$ & $1.3665591366778236*10^{-9}$ \\ 
\hline 
1000 & $2.8318599999999999*10^{-8}$ & $1.2139629791575567*10^{-9}$ \\ 
\hline
\end{tabular} 
\caption{Values of $M_\Delta$, $\mu$ and $v_\Delta$ allowed by the Eq. (\ref{ISSequation}) that we used in our analysis of the LHC.}
\label{table3}
\end{table}
\begin{figure}[tb]
\begin{center}
	\includegraphics[scale=0.17]{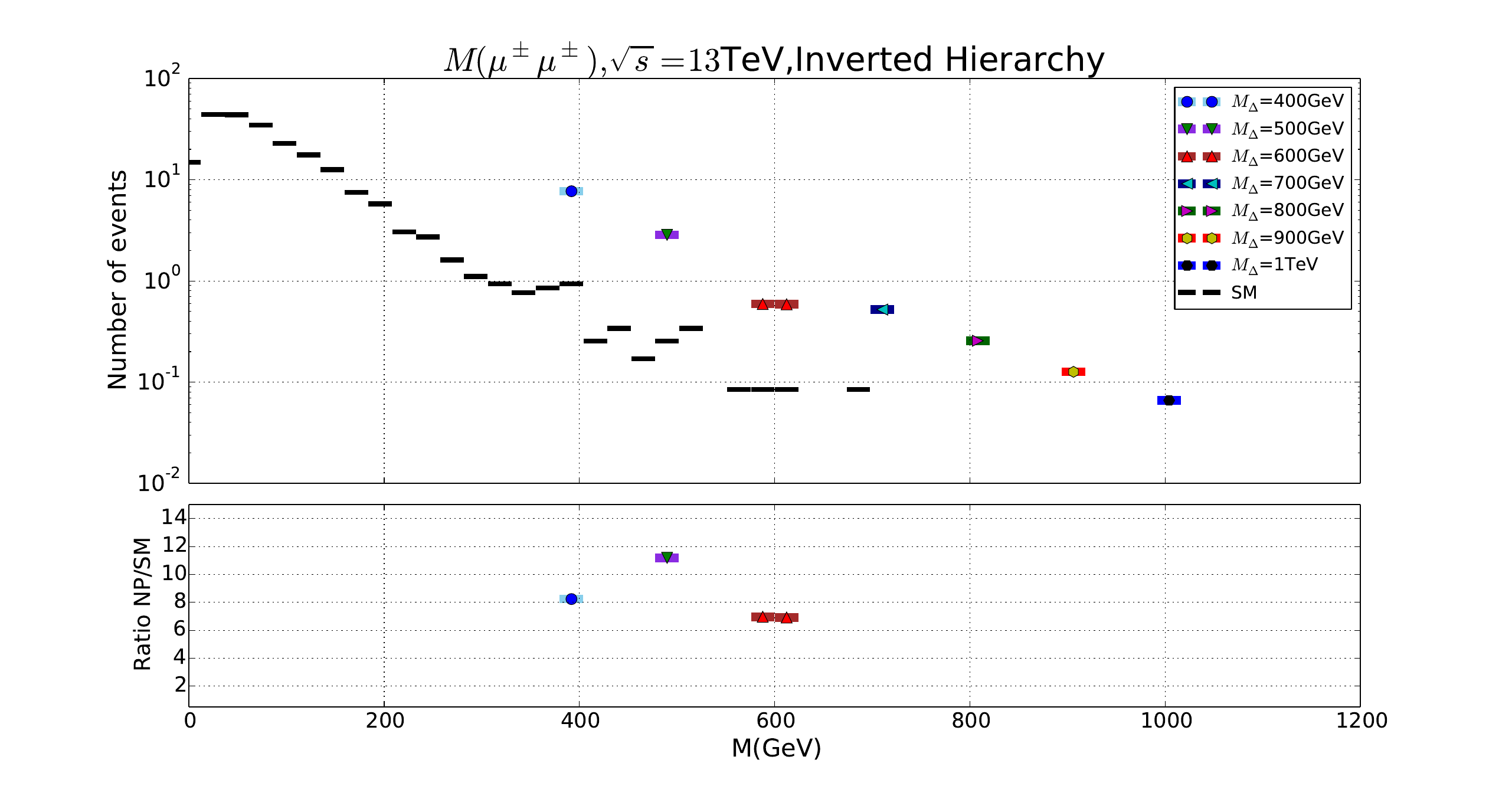} \quad
	\includegraphics[scale=0.17]{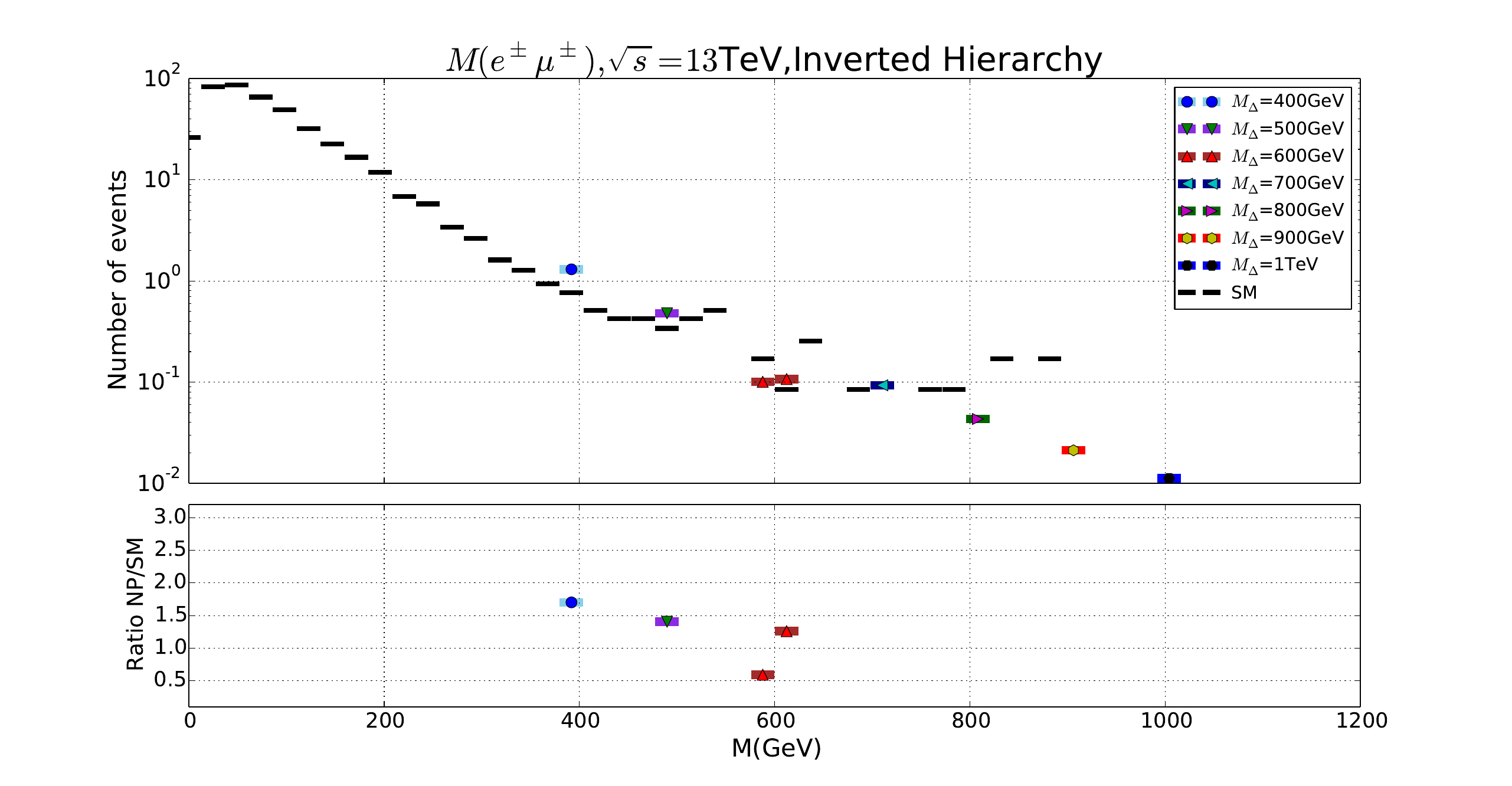} \quad
	\includegraphics[scale=0.20]{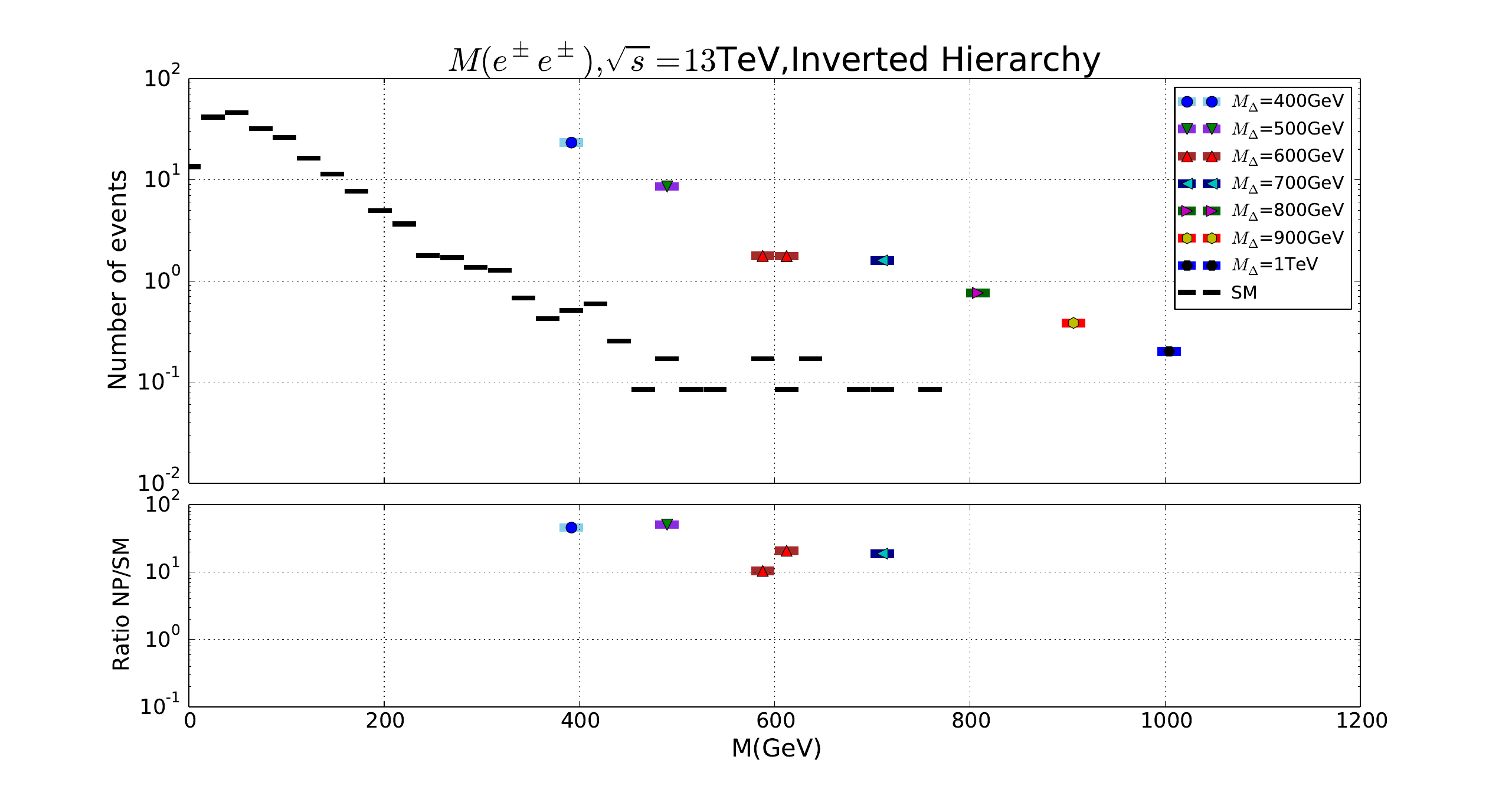}
\caption{The process $q + \bar q \rightarrow \delta^{++} +\delta^{--} \rightarrow l^+ l^- l^+ l^-$  (with $l=e, \mu$) at the LHC  for the case of IH with  $\sqrt{s}=13$TeV.}
 \label{fig8}
\end{center}
\end{figure}

\section{Summary and Conclusions}
\label{sec7}

In this work we revisited and  probed the signature of what we called  the inverse type II seesaw mechanism. Such mechanism was implemented in the framework of  low scale Higgs triplet model and its may signature are new Higgs fields that may be probed at the LHC and ILC. In this paper we restricted our analysis to  the production of doubly charged scalars, only. Such scalars will have, as  distinctive   properties, mass in the range of hundreds of GeVs until TeV and couplings with leptons (Yukawa couplings)  fixed by  neutrino masses and mixing angles in accordance to recent measurements of neutrino oscillations.

At the ILC we probed  $\delta^{++}$ through the process $e^+ + e^- \rightarrow \delta^{++} +\delta^{--} \rightarrow l^+ l^- l^+ l^-$ with $l=e, \mu$. Our results were presented in plots relating the number of events to the invariant mass of pairs of charged leptons.  We also presented plots comparing the strength of the  signal due to the new physics with  the background due to the standard model.

We considered ILC running from 500 until 1TeV. In both situations the ILC presented high efficiency   in producing   events generated by this process.  However, because the ILC is restricted to energy scale of at most 1TeV, then only doubly charged scalar with mass at most of 500 GeV may be probed in such a machine. It is also important to say that the ILC is not the best place to distinguish the NH case from the IH one.   As result we say that the ILC is the fairest place to find such scalars once robust number of events are produced with  large signals. 

At the LHC things are completely different. In it the doubly charged scalars are probed through the process $pp  \rightarrow \delta^{++} +\delta^{--} \rightarrow l^+ l^- l^+ l^-$ with $l=e, \mu$.  Our analysis were done for $\sqrt{s}=13$ TeV and luminosity of $41,07$fb$^{-1}$.  The LHC is not efficient when NH case is considered, but presents sizable, but not  comparable to the ILC, efficiency in producing such  doubly charged scalars in the case of  IH scenario. Thus,  if some  electrons and positrons are found as final product of these processes,  then we may say that they are  results of the inverted hierarchy scenario. In other word, the LHC may help in distinguishing  the hierarchy of the neutrino masses. Thus, we conclude that such machines, LHC and ILC, complement each other in the search for the signature of the inverse seesaw mechanism promoted by the low energy scale Higgs triplet model.

\begin{acknowledgments}
FFF is supported by Coordena\c{c}\~ao de Aperfei\c{c}oamento de Pessoal de N\'{\i}vel Superior (CAPES). CASP and PSRS are supported by Conselho Nacional de Pesquisa e Desenvolvimento 
Cient\'{\i}fico - CNPq. The authors thanks Alexandre Alves and J. G. J\'unior for useful discussions.
\end{acknowledgments}


\begin{thebibliography}{}
\bibitem{note}
We have to remark that, as we saw in the end of Sec.~\ref{sec5}, there is a $\delta^{++}$ decay channel involving the $\tau$ lepton which is important too. However, since the hadronically decaying $\tau$ is harder to measure and is not as clean as the lighter leptonic decays, we leave it outside this analysis.
\end{thebibliography}
\end{document}